\documentclass[aps,prl,citeautoscript,reprint,superscriptaddress,floatfix,footinbib,showkeys]{revtex4-2}
\usepackage{setspace}
\usepackage[utf8]{inputenc}
\usepackage[colorlinks=true,linkcolor=black,urlcolor=black,filecolor=black,citecolor=black]{hyperref}
\usepackage{color,soul}
\usepackage{amsmath}
\usepackage{amsfonts}
\usepackage{amssymb}
\usepackage{graphicx}
\usepackage{dcolumn}
\usepackage{bm}
\usepackage{array}
\usepackage{booktabs}

\begin{document}
\title{Ballistic-to-diffusive transition in engineered counter-propagating quantum Hall channels}
\author{Aifei Zhang}
\affiliation{Université Paris-Saclay, CEA, CNRS, SPEC, 91191 Gif-sur-Yvette, France} 
\author{K. Watanabe}
\affiliation{Research Center for Electronic and Optical Materials, National Institute for Materials Science, 1-1 Namiki, Tsukuba 305-0044, Japan} 
\author{T. Taniguchi} 
\affiliation{Research Center for Materials Nanoarchitectonics, National Institute for Materials Science,  1-1 Namiki, Tsukuba 305-0044, Japan}
\author{Patrice Roche} 
\affiliation{Université Paris-Saclay, CEA, CNRS, SPEC, 91191 Gif-sur-Yvette, France} \author{Carles Altimiras} \affiliation{Université Paris-Saclay, CEA, CNRS, SPEC, 91191 Gif-sur-Yvette, France} \author{François D. Parmentier} \affiliation{Université Paris-Saclay, CEA, CNRS, SPEC, 91191 Gif-sur-Yvette, France} \affiliation{Laboratoire de Physique de l’Ecole normale sup\'erieure, ENS, Universit\'e PSL, CNRS, Sorbonne Universit\'e, Universit\'e Paris Cit\'e, F-75005 Paris, France } \author{Olivier Maillet}\email{olivier.maillet@cea.fr}
\affiliation{Université Paris-Saclay, CEA, CNRS, SPEC, 91191 Gif-sur-Yvette, France}


\begin{abstract}
Exotic quantum Hall systems hosting counter-propagating edge states can show seemingly non-universal transport regimes, usually depending on the size of the sample. We experimentally probe transport in a quantum Hall sample engineered to host a tunable number of counter-propagating edge states. The latter are coupled by Landauer reservoirs, which force charge equilibration over a tunable effective length. We show that charge transport is determined by the balance of up- and downstream channels, with a ballistic regime emerging for unequal numbers of channels. For equal numbers, we observe a transition to a critical diffusive regime, characterized by a diverging equilibration length. Our approach allows simulating the equilibration of hole-conjugate states and other exotic quantum Hall effects with fully controlled parameters using well-understood quantum Hall states.\end{abstract} 

\maketitle
Quantum Hall (QH) insulators are characterized by a gapped, electrically insulating bulk, and conduction along their edge via a quantized number of one-dimensional chiral edge channels. In most cases, the edge channels co-propagate, \textit{i.e.} they have the same chirality. This leads to extremely robust transport properties, even in presence of strong interactions between edge channels \cite{Neder2006PRL,leSueur2010PRL}: edge transport is then ballistic and dissipationless, and the Hall conductance is exactly given by $\nu e^2/h$, where $e$ is the electron charge, $h$ Planck’s constant, and $\nu=nh/eB$ is the filling factor characterizing the topology of the QH states ($n$ is the carrier density and $B$ the perpendicular magnetic field). However, the edge channels can also be counter-propagating; this is notably the case of the quantum spin Hall (QSH) effect, which hosts two counterpropagating channels with opposite spin polarizations~\cite{Konig2007SciSQH}, but it is also expected for hole-conjugate states of the fractional quantum Hall (FQH) effect such as $\nu=2/3$ \cite{McDo_fractional_edges_1990}. There, Coulomb interactions and inter-channel tunneling can lead to equilibration along the edge \cite{KFP_equilibration_1994}, drastically impacting transport properties, which become length-dependent. Theoretical \cite{ProtopopovAnnals2017,NosigliaPRB2018,ParkPRB2019,SpanslattPRL2019,SpanslattPRB2020,FujisawaPRB2021,Glattli2024arxiv} and experimental \cite{Bid2010,Lin2019PRB,Cohen2019NComms,Lafont2019Science,SrivastavPRL2021,Melcer2022NatComms} investigations of charge and heat transport in hole-conjugate FQH states, combined with the general observation of imperfect conductance quantization in the QSH effect \cite{Konig2007SciSQH,Roth2009Science,Knez2011,Wu2018QSH,Veyrat2020Science,Kang2024,Yang2024NComms}, point towards different transport regimes depending on the number of equilibrating edge channels and their respective conductances. If the overall upstream and downstream conductances are different, the transport can be ballistic, with a quantized conductance and negligible dissipation at large scale. If upstream and downstream conductances are equal, however, the transport becomes diffusive, with inexact conductance quantization and dissipation all along the edge. Remarkably, this applies to both charge and heat transport, which can lead to striking transport decoupling effects in FQH channels where the electrical conductance is fractional while the thermal conductance is integer~\cite{Banerjee2017Nature, SrivastavPRL2021, Melcer2022NatComms}.
 
Understanding and exploring the transition between the diffusive and ballistic regimes in equilibrating counterpropagating edge channels is an experimental challenge, as it requires controlling the number of equilibrating channels, the nature and strength of the equilibration process, as well as the ability to probe dissipation along the edge of the sample. Inspired by recent theoretical models~\cite{SpanslattPRL2019,SpanslattPRB2020, Glattli2024arxiv}, we have realized an experiment where counterpropagating integer QH channels equilibrate through a series of Ohmic contacts (Landauer reservoirs) which provide both charge redistribution and energy equilibration between channels. Our approach presents multiple advantages, stemming from its simplicity: for any edge configuration among a wide range, we directly measure the voltage of the Ohmic contacts, yielding the voltage drop and dissipation along the edge, as well as the conductance and its length dependence. We emulate charge equilibration in a QH edge with tunable numbers of counter-propagating channels. This allows us to observe and fully explore the transition between the ballistic and diffusive regimes, in agreement with exact scattering formalism calculations.


\begin{figure*}[ht!]
	\centering
	\includegraphics[width=1\textwidth]{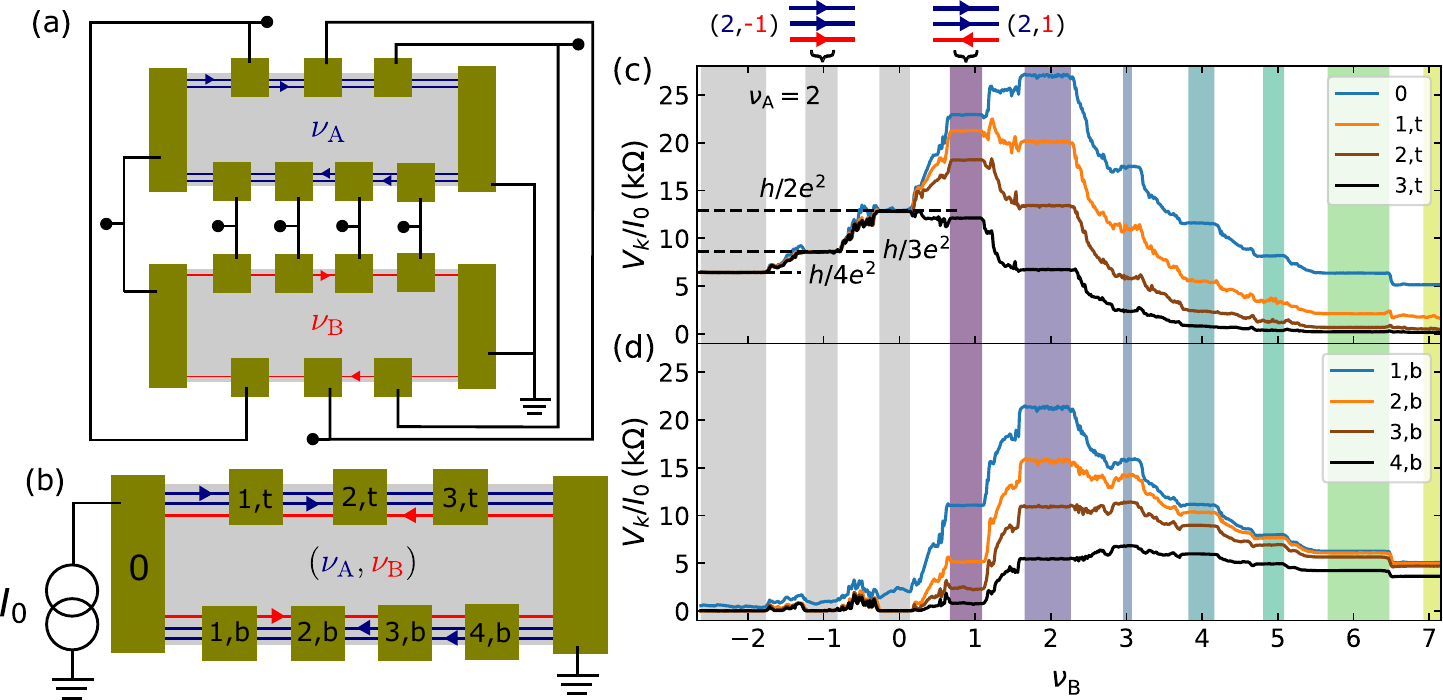}
	\caption{a. Schematics of the measured two-Hall bar device and paired contacts. b. Effective Hall bar: equivalent configuration with counter-propagating edge channel numbers set by individual filling factors $\nu_\mathrm{A}$ and $\nu_\mathrm{B}$. In this example, one pair of contacts (labeled 0 here) is connected to a current source (injection contact), and $N_\mathrm{t}=3$ and $N_\mathrm{b}=4$ intermediate contacts on the top and bottom edge, respectively, are left floating, acting as effective Landauer reservoirs. c-d. Example sweep of device B's gate voltage/density (in filling factor units $\nu_\mathrm{B}$) while keeping $\nu_\mathrm{A}=2$, for top (panel c) and bottom edge contacts (panel d). The colors assigned to each plateau with $\nu_\mathrm{B}>0$ are used throughout the letter.}\label{fig_sweep}
\end{figure*}

Our experiment consists of two Hall bars (device A and B, see Fig. \ref{fig_sweep}a) made of monolayer graphene encapsulated in hexagonal boron nitride \cite{Dean2010NatNano}, each individually gated by a graphite back gate. Both are cooled down to 10 mK under a 14 T perpendicular magnetic field. The Hall bars are connected in a top-to-tail fashion, such that the $k^\mathrm{th}$ contact of device A, starting clockwise from the first contact next to the cold-grounded drain, is connected with the $k^\mathrm{th}$ contact of device B, starting anticlockwise from the first contact next to the drain. Paired contacts share the same potential, and each pair may be left floating (while its voltage with respect to ground is measured), or put to ground, or used for current injection. This implements an effective single Hall bar (Fig. \ref{fig_sweep}b) with a tunable number of counter-propagating channels determined by the filling factor of each Hall bar $\nu_\mathrm{A/B}$. By convention, we define the edge flowing from (resp. into) the injection contact, when running clockwise on the effective Hall bar, as the top (resp. bottom) edge, labeled ``t" (resp. ``b"). We introduce $N_\mathrm{t}$ (resp. $N_\mathrm{b}$) the number of intermediate floating Ohmic contacts on the top (resp. bottom) edge. The gate-to-density correspondence of one device is determined by setting the other device at $\nu=0$, thus ensuring that current flows only in the former \cite{SM}. We then keep device A at a fixed filling factor $\nu_\mathrm{A}$, and tune the charge carrier density $n_\mathrm{B}$ of device B to obtain filling factors ranging from $\nu_\mathrm{B}=-2$ to $\nu_\mathrm{B}=7$. 
We inject a low frequency ($\approx 1$ Hz) current through the source contacts of our choice in both devices, and monitor voltages on each pair of contacts, including the source, through standard lock-in measurements. A sweep of the back gate of device B for $\nu_\mathrm{A}=2$ in the configuration of Fig. \ref{fig_sweep}b) is shown in Figure \ref{fig_sweep}c-d). When device B is p-doped ($\nu_\mathrm{B}<0$), the voltage on the top effective edge (``t"-labeled contacts in Fig. \ref{fig_sweep}c) remains constant and equal to that of the injection, while voltages on all bottom edge contacts remain zero, up to some residual back-scattering observable on contact 1,b. This is expected since here, all channels co-propagate, making our effective Hall bar behave like a regular one. 
On the contrary, for $\nu_\mathrm{B}>0$, we observe a voltage decrease on the top edge starting from the source contact, while we measure non-zero voltages on the bottom edge contacts, increasing towards the source. In that configuration, from the effective Hall bar's perspective, edge channels from devices A and  B have opposite chiralities, enabling equilibration in the intermediate contacts.
\begin{figure*}
	\centering
\includegraphics[width=1\textwidth]{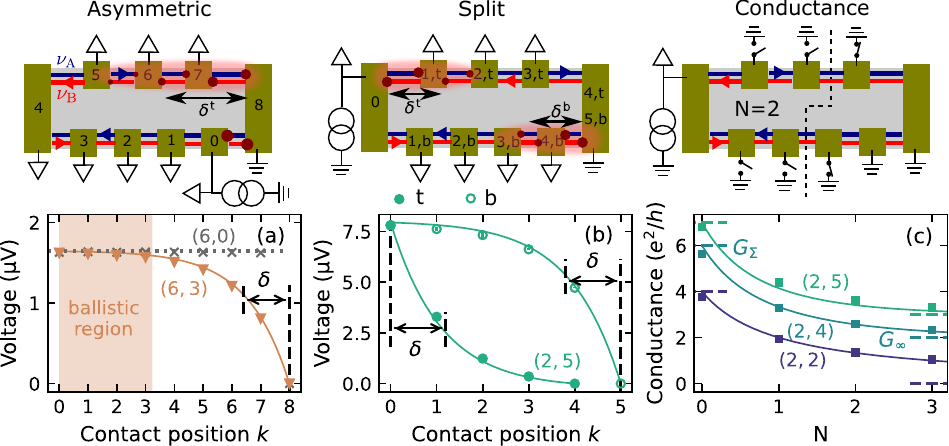}
	\caption{a) Sketch of the asymmetric configuration and corresponding voltage drop $V_k$ measurements for $\nu_\mathrm{A}=6$, and $\nu_\mathrm{B}=0$ (gray crosses) and $3$ (orange triangles), for an injected current amplitude $I_0\approx 0.38$ nA. b) Sketch of the split configuration and corresponding voltage drop measurements or $\nu_\mathrm{A}=2$ and $\nu_\mathrm{B}=5$, with injected current amplitude $I _0\approx 0.96$ nA. Open (resp. full) dots correspond to bottom (resp. top) edge contacts. Solid lines in both panel are applications of Eq. (\ref{Vprofile_top_edge}) for corresponding edges and bulk filling factors, and the dotted line in a) is a guide for the eye at constant $V_k$. c) Two-point conductance configuration and measurements (square symbols) for ($\nu_\mathrm{A}=2,\nu_\mathrm{B}=2,4,5$), obtained when adding pairs of grounds on each side, while keeping the bottom rightmost contact grounded to keep a symmetric configuration. Solid lines are applications of Eq. (\ref{G_2pt_LB}). Dashed lines represent the limit values of zero ($G_\Sigma$) and full ($G_\infty$) equilibration.}
    \label{fig_raw}
\end{figure*}

These observations can be all captured with a simple approach for chiral edge currents: the current emitted in one IQH channel from contact $k$ is $e^2V_k/h$. We assume that ideal equilibration occurs in each intermediate metallic contact and that propagation in between contacts, beyond negligible bulk leakage, is fully ballistic, insofar as it does not suffer from local (\textit{i.e.} microscopic) equilibration. This last assumption is well satisfied in practice, because of the physical separation between channels of opposite chirality. Current conservation at contact $k,\mathrm{t}$ leads to $(\nu_\mathrm{A}+\nu_\mathrm{B})V_{k,\mathrm{t}}=\nu_\mathrm{A}V_{k-1,\mathrm{t}}+\nu_\mathrm{B}V_{k+1,\mathrm{t}}$, with $V_{N+1}=0$ for the drain contact. We obtain the potential at each contact $k$ on the top edge when $\nu_\mathrm{A}\neq\nu_\mathrm{B}$:
\begin{equation}
    \label{Vprofile_top_edge}
    V_{k,\mathrm{t}}=\frac{1-\exp{\left[-(N_\mathrm{t}+1-k)\log\left(\nu_\textrm{A}/\nu_\textrm{B}\right)\right]}}{1-\exp{\left[-(N_\mathrm{t}+1)\log\left(\nu_\textrm{A}/\nu_\textrm{B}\right)\right]}}V_0,
\end{equation}
with $V_0$ the source voltage \cite{SM}. The same result is derived for voltages on the $N_\mathrm{b}$ bottom edge contacts, up to a swap between $\nu_\mathrm{A}$ and $\nu_\mathrm{B}$. According to Eq.~(\ref{Vprofile_top_edge}), for $\nu_\mathrm{B}\neq\nu_\mathrm{A}$, the voltage drop along the edge presents an exponential profile, highlighting a characteristic dimensionless equilibration distance $\delta=1/|\log(\nu_\mathrm{A}/\nu_\mathrm{B})|$ that should be compared with the effective edge length $N_\mathrm{t}$.


This exponential profile is clearly observed in the data: Fig.~\ref{fig_raw} shows voltages measured at the source and all floating contacts, for two different measurement configurations. In the first ``asymmetric" configuration (Fig.~ \ref{fig_raw}a), all the contacts are located on the top effective edge ($N_\mathrm{t}=7,N_\mathrm{b}=0$), so that equilibration only occurs on the top edge, and can be probed over a large number of contacts. For $(\nu_\mathrm{A},\nu_\mathrm{B})=(6,0)$ (no counter-propagating channel), the voltage stays constant over the whole top edge up to the drain. For $\nu_\mathrm{B}>0$, e.g. the $(6,3)$ case represented in Fig. \ref{fig_raw}a), the voltage remains constant and equal to that of the source contact, until a drop occurs within the last few contacts before the drain. This drop corresponds to energy dissipation over a portion of the edge, highlighted as a ``hot spot" in red in Fig. \ref{fig_raw}a)-b) schematics. The measured voltage profile is exactly matched by Eq.~(\ref{Vprofile_top_edge}), particularly the exponential dependence parametrized by the dimensionless length $\delta$.

In the second, ``split" configuration, $N_\mathrm{t}=3$ contacts are located on the top edge, and $N_\mathrm{b}=4$ on the bottom edge, allowing probing the two edges independently. Fig. \ref{fig_raw}b) shows both voltage profiles for the $(2,5)$ case (where $\nu_\mathrm{A}<\nu_\mathrm{B}$), in excellent agreement with Eq.~\ref{Vprofile_top_edge}. On the top edge (where more channels come from the drain at zero potential than from the injection contact), the voltage rapidly drops to zero over the expected length $\delta$. Contrarily, on the bottom edge, the voltage remains close to $V_0$, only dropping close the drain over the same length $\delta$. Therefore, the voltage profile on a given edge (saturation value and drop region) is dictated by its dominant chirality. Furthermore, it signals a ballistic behavior outside of the voltage drop region, when the effective edge length (the number of contacts) exceeds a few $\delta$, as shown in Fig.~\ref{fig_raw}a). This is qualitatively reminiscent of dissipation in a standard QH system, where the voltage along a given chiral edge channel is constant, and only drops at the downstream contact, with a fully localized hot spot. Thus, we use hereafter the term ``downstream" to indicate the end contact (source or drain) closer to which the voltage drop occurs for a given edge. We also denote $\Delta V_\mathrm{end}$ the voltage drop between the last floating contact and the downstream contact. 

In addition to the voltage profile measurements, we probe the length dependence of the two point conductance of the effective Hall bar. In this configuration (see Fig.~\ref{fig_raw}c), the bottom contact closest to the drain is grounded to obtain a fully symmetric device. We change the sample's effective length by successively grounding pairs of contacts facing each other, starting from the drain side, thus leaving $N$ pairs of floating contacts between the source and the grounded contacts. $N=0$ corresponds to all contacts grounded apart from the injection one (and thus no equilibration), while $N=3$ corresponds to none grounded except the bottom one closest to drain. The conductance in this configuration can be calculated from Eq.~(\ref{Vprofile_top_edge}), for $\nu_\mathrm{A}\neq\nu_\mathrm{B}$:
\begin{equation}
    \label{G_2pt_LB}
    G_\mathrm{2w}=G_\infty\coth{\left[\frac{N+1}{2}\left|\log(\nu_\mathrm{A}/\nu_\mathrm{B})\right|\right]}.
\end{equation} 

This expression also involves the characteristic length $\delta=1/|\log(\nu_\mathrm{A}/\nu_\mathrm{B})|$, and shows an exponential convergence to the equilibrated value $G_\infty=|\nu_\mathrm{A}-\nu_\mathrm{B}|e^2/h$ for large samples (that is, $N\gg \delta$).
\begin{figure}
	\centering
\includegraphics[width=0.48\textwidth]{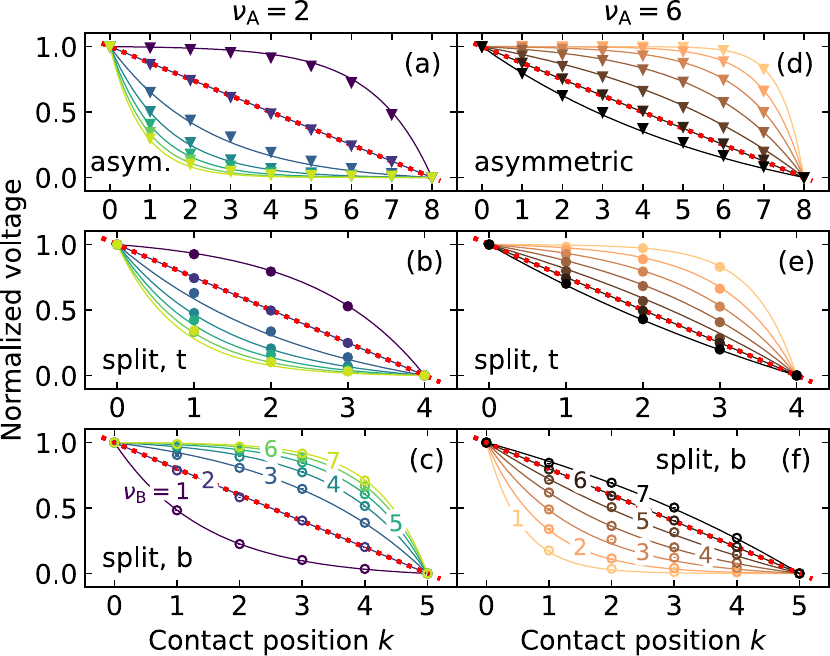}
	\caption{Normalized voltage profiles $V_k/V_0$ for all $\nu_\mathrm{B}$ values at $\nu_\mathrm{A}=2$ (a,b,c) and 6 (d,e,f). Solid lines are derived from the scattering approach. Linear profiles are singled out for $\nu_\mathrm{A}=\nu_\mathrm{B}$ (red dotted lines).}\label{fig_trans}
\end{figure}

The measured conductances are displayed in Fig. \ref{fig_raw}c) for $\nu_\mathrm{A}=2$ and $\nu_\mathrm{B}=2,4,5$, showing an excellent agreement with Eq.~\ref{G_2pt_LB}. We observe a clear decrease of conductance with the number $N$ of intermediate contacts, from a value close to $G_\Sigma=(\nu_\mathrm{A}+\nu_\mathrm{B})e^2/h$, which corresponds to decoupled channels at $N=0$. Conductances are close to $G_\infty$ already at $N=3$. For the singular case $\nu_\mathrm{A}=\nu_\mathrm{B}$, the conductance slowly decreases and remains substantially above $G_\infty=0$ at $N=3$, and is well matched by the corresponding formula
$G_\mathrm{2w}= G_\Sigma/(N+1)$ (see \cite{SM}).

The divergence of $\delta$ translates as a voltage profile that approaches the linear trend as $\nu_\mathrm{B}\rightarrow\nu_\mathrm{A}$, and becomes linear for the singular case $\nu_\mathrm{B}=\nu_\mathrm{A}$: $V_{k,\gamma=\mathrm{t,b}}=V_0(N_\gamma+1-k)/(N_\gamma+1)$, as shown in Fig. \ref{fig_trans}a-f). This corresponds to a transition to a diffusive, Ohmic regime, with a hot spot that is delocalized over the whole edge, since all voltage drops between successive contacts have the same value $-V_0/(N_\gamma+1)$ independent of their position. This also appears in the conductance: for $\nu_\mathrm{A}=\nu_\mathrm{B}$, we have $G_\mathrm{2w}=2\nu_\mathrm{A}e^2/(N+1)h$, which is confirmed experimentally through the slow, algebraic convergence of the (2,2) conductance to zero (Fig. 2c).
\begin{figure}
	\includegraphics[width=0.48\textwidth]{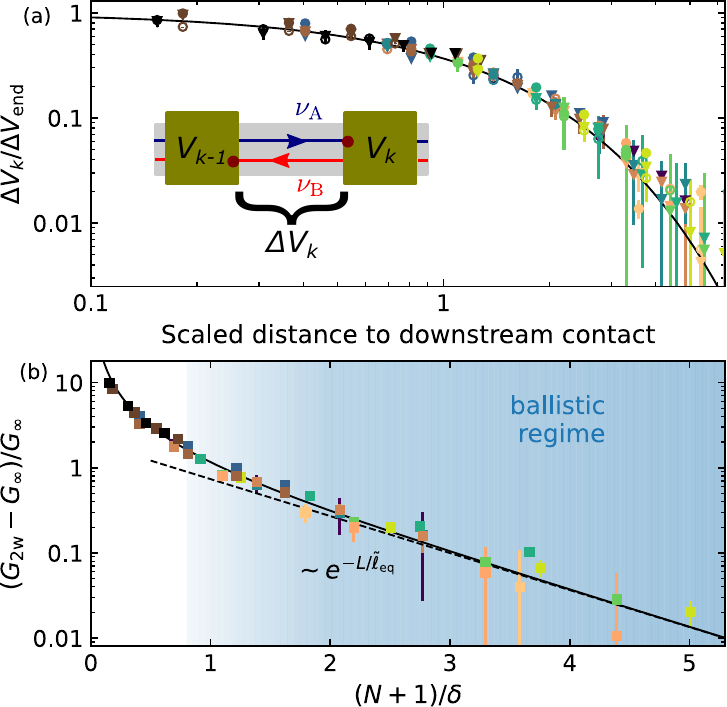}
	\caption{(a) Voltage drops between successive contacts, normalized to the maximum voltage drop $\Delta V_\mathrm{end}$, as a function of $\lambda/\delta$, the distance to the downstream contact normalized to the equilibration distance (see text). Solid line: $e^{-\lambda/\delta}$. (b) Normalized two-point conductance of the effective Hall bar in the ``conductance" configuration, as a function of the scaled effective bar's length. Solid line: application of Eq. (\ref{G_2pt_LB}). The dashed line corresponds to the asymptotic limit of a ballistic ($L\gg\tilde{\ell}_\mathrm{eq}$) regime, with $N$ identified to $L/\ell_\mathrm{eq}$.}\label{fig_univ}
\end{figure}

Our results highlight universal behaviors for both voltage profiles and conductance. For $\nu_\mathrm{A}\neq\nu_\mathrm{B}$, the potential difference between counter-propagating channels is given by the voltage difference between successive contacts, which can be written, following Eq. (\ref{Vprofile_top_edge}), $\Delta V_{k,\gamma}\equiv V_{k,\gamma}-V_{k-1,\gamma}=e^{-\lambda/\delta}\Delta V_\mathrm{end,\gamma}$ ($\lambda=k$ or $N_\gamma+1-k$ depending on the dominant chirality on the considered edge). Therefore, equilibration is ensured for distances to downstream contact $\lambda\gg\delta$, where $\Delta V_k\approx 0$. When $\lambda$ is rescaled to the equilibration distance $\delta$, all voltage drops should collapse on a single curve and fall exponentially to zero. This is shown clearly in Fig. \ref{fig_univ}a) for all the data presented in Fig. \ref{fig_trans} (colors correspond to the filling factor combinations of Fig. \ref{fig_trans}). 

This rescaling suggests a correspondence between our experiment and a sample with counter-propagating channels experiencing charge equilibration over a continuous edge \cite{Lin2019PRB,Lafont2019Science,Cohen2019NComms}. Indeed the number of Landauer reservoirs may be mapped to a continuous edge's length in units of is bare equilibration length, $N\equiv L/\ell_\mathrm{eq}$. This analogy is strengthened by the similarity between Eq. (\ref{G_2pt_LB}) and the conductance of a continuous edge sample as a function of its length $L$ in the limit of $L\gg\ell_\mathrm{eq}$ \cite{ProtopopovAnnals2017,SrivastavPRL2021}. As shown in Fig. \ref{fig_univ}b), all reduced conductance data (see \cite{SM} for the full conductance data sets for $\nu_\mathrm{A}=(2,6)$ and $\nu_\mathrm{B}=1 \leftrightarrow 7$) $(G_\mathrm{2w}-G_\infty)/G_\infty$ collapse on a single curve $\coth(x/2)-1$, with $x=(N+1)/\delta$. For an effectively large sample, \textit{i.e.} $N\gg\delta$, $G_{2\mathrm{w}}$ converges exponentially towards its equilibrated value $G_\infty$, \textit{i.e.} $(G_{2\mathrm{w}}-G_\infty)/G_\infty\sim e^{-N/\delta}\equiv e^{-L/\tilde{\ell}_\mathrm{eq}}$. Here we have operated a discrete-to-continuum correspondence, with $\tilde{\ell}_\mathrm{eq}=\ell_\mathrm{eq}/|\log(\nu_\mathrm{A}/\nu_\mathrm{B})|$ the \textit{effective} equilibration length which absorbs the contribution from filling factors measured in our experiment.

We now discuss the relation with equilibration models used for hole-conjugate states \cite{KFP_equilibration_1994,ProtopopovAnnals2017,Lin2019PRB,FujisawaPRB2021,SpanslattPRL2019}, particularly $\nu=2/3$, in which charge transport exemplifies the $\nu_\mathrm{A}\neq\nu_\mathrm{B}$ case. In these models, the potential difference between counter-propagating edges drives inter-channel charge tunneling, leading to equilibration over an effective length $\ell_\mathrm{eq}/|\nu_\mathrm{A}^{-1}-\nu_\mathrm{B}^{-1}|$ \cite{Lin2019PRB,SpanslattPRB2020}, with $\ell_\mathrm{eq}$ the bare equilibration length that is usually obtained in the Lüttinger liquid framework \cite{ProtopopovAnnals2017,SrivastavPRL2021}. Our definition of $\tilde{\ell}_\mathrm{eq}$ differs from this one, because the fundamentally different equilibration mechanisms at play (inter-edge tunneling vs reservoir redistribution). However, in the critical regime $\nu_\mathrm{A}\rightarrow\nu_\mathrm{B}$, both definitions lead to the same divergence $\delta\sim1/|\nu_\mathrm{A}-\nu_\mathrm{B}|$. In the diffusive case $\nu_\mathrm{A}=\nu_\mathrm{B}$, the two point conductance in the macroscopic limit $L\gg\ell_\mathrm{eq}$ behaves according to a universal Ohmic scaling law, decaying algebraically: $G_\mathrm{2w}\propto 1/L$ (see Fig. \ref{fig_raw}c), irrespective of the microscopic mechanism of equilibration \cite{KFP_equilibration_1994,FujisawaPRB2021,Glattli2024arxiv,SM}. This is reminiscent of the Ohmic deviations to quantized transport observed in QSH samples \cite{Wu2018QSH}, where the resistance scales roughly linearly with the sample's length for large samples, and analogous to the transport of heat for the disordered $\nu=2/3$ edge, where thermal conductances are equal for charged and neutral modes, leading to diffusive signatures in noise while charge transport remains ballistic
\cite{KaneFisherPRB95,NosigliaPRB2018,SpanslattPRL2019,SpanslattPRB2020,SrivastavPRL2021,Melcer2022NatComms,HashisakaPRX2023}. In that respect, $\nu=2/3$ is formally equivalent to the cases (2,6) and (6,2) that we study for charge, and would be equivalent to (2,2) or (6,6) for heat transport.


In conclusion, our experiment explores the transition between an effective ballistic edge transport (despite the presence of counter-propagating edge states) and a critical, scale-invariant diffusive regime. It conceptual simplicity makes it applicable to a large variety of systems with counterpropagating edge channels: beyond hole-conjugate fractional and QSH states, it also encompasses the recent observation of robust quantization to the equilibrated value for the valley-polarized $\nu=-2$ state in PbSnSe Dirac systems~\cite{KrizmanPRL2024}
A possible extension of this work is the investigation of the heat flow in such systems via noise measurements, e.g. to witness charge-heat separation with several fractional states, or when adding interactions, as in the case of Heat Coulomb Blockade \cite{Sivre2018,Stabler2023PRB,stabler2024giantheatfluxeffect}.

\section{Acknowledgments}
We thank D. C. Glattli, P. Joyez and C. Mora for helpful discussions. This work was funded by the ERC (ERC-2018-STG \textit{QUAHQ}), by the “Investissements d’Avenir” LabEx PALM (ANR-10-LABX-0039-PALM), and by the Region Ile de France through the DIM QUANTIP. O.M. acknowledges funding from the ANR (ANR-23-CE47-0002 CRAQUANT). K.W. and T.T. acknowledge support from the JSPS KAKENHI (Grant Numbers 21H05233 and 23H02052) , the CREST (JPMJCR24A5), JST and World Premier International Research Center Initiative (WPI), MEXT, Japan. 

\section{Competing interests}

The authors declare no competing interests.

\bibliography{KFP_simu}

\begin{thebibliography}{34}%
\makeatletter
\providecommand \@ifxundefined [1]{%
 \@ifx{#1\undefined}
}%
\providecommand \@ifnum [1]{%
 \ifnum #1\expandafter \@firstoftwo
 \else \expandafter \@secondoftwo
 \fi
}%
\providecommand \@ifx [1]{%
 \ifx #1\expandafter \@firstoftwo
 \else \expandafter \@secondoftwo
 \fi
}%
\providecommand \natexlab [1]{#1}%
\providecommand \enquote  [1]{``#1''}%
\providecommand \bibnamefont  [1]{#1}%
\providecommand \bibfnamefont [1]{#1}%
\providecommand \citenamefont [1]{#1}%
\providecommand \href@noop [0]{\@secondoftwo}%
\providecommand \href [0]{\begingroup \@sanitize@url \@href}%
\providecommand \@href[1]{\@@startlink{#1}\@@href}%
\providecommand \@@href[1]{\endgroup#1\@@endlink}%
\providecommand \@sanitize@url [0]{\catcode `\\12\catcode `\$12\catcode `\&12\catcode `\#12\catcode `\^12\catcode `\_12\catcode `\%12\relax}%
\providecommand \@@startlink[1]{}%
\providecommand \@@endlink[0]{}%
\providecommand \url  [0]{\begingroup\@sanitize@url \@url }%
\providecommand \@url [1]{\endgroup\@href {#1}{\urlprefix }}%
\providecommand \urlprefix  [0]{URL }%
\providecommand \Eprint [0]{\href }%
\providecommand \doibase [0]{https://doi.org/}%
\providecommand \selectlanguage [0]{\@gobble}%
\providecommand \bibinfo  [0]{\@secondoftwo}%
\providecommand \bibfield  [0]{\@secondoftwo}%
\providecommand \translation [1]{[#1]}%
\providecommand \BibitemOpen [0]{}%
\providecommand \bibitemStop [0]{}%
\providecommand \bibitemNoStop [0]{.\EOS\space}%
\providecommand \EOS [0]{\spacefactor3000\relax}%
\providecommand \BibitemShut  [1]{\csname bibitem#1\endcsname}%
\let\auto@bib@innerbib\@empty
\bibitem [{\citenamefont {Neder}\ \emph {et~al.}(2006)\citenamefont {Neder}, \citenamefont {Heiblum}, \citenamefont {Levinson}, \citenamefont {Mahalu},\ and\ \citenamefont {Umansky}}]{Neder2006PRL}%
  \BibitemOpen
  \bibfield  {author} {\bibinfo {author} {\bibfnamefont {I.}~\bibnamefont {Neder}}, \bibinfo {author} {\bibfnamefont {M.}~\bibnamefont {Heiblum}}, \bibinfo {author} {\bibfnamefont {Y.}~\bibnamefont {Levinson}}, \bibinfo {author} {\bibfnamefont {D.}~\bibnamefont {Mahalu}},\ and\ \bibinfo {author} {\bibfnamefont {V.}~\bibnamefont {Umansky}},\ }\bibfield  {title} {\bibinfo {title} {Unexpected behavior in a two-path electron interferometer},\ }\href {https://doi.org/10.1103/PhysRevLett.96.016804} {\bibfield  {journal} {\bibinfo  {journal} {Phys. Rev. Lett.}\ }\textbf {\bibinfo {volume} {96}},\ \bibinfo {pages} {016804} (\bibinfo {year} {2006})}\BibitemShut {NoStop}%
\bibitem [{\citenamefont {le~Sueur}\ \emph {et~al.}(2010)\citenamefont {le~Sueur}, \citenamefont {Altimiras}, \citenamefont {Gennser}, \citenamefont {Cavanna}, \citenamefont {Mailly},\ and\ \citenamefont {Pierre}}]{leSueur2010PRL}%
  \BibitemOpen
  \bibfield  {author} {\bibinfo {author} {\bibfnamefont {H.}~\bibnamefont {le~Sueur}}, \bibinfo {author} {\bibfnamefont {C.}~\bibnamefont {Altimiras}}, \bibinfo {author} {\bibfnamefont {U.}~\bibnamefont {Gennser}}, \bibinfo {author} {\bibfnamefont {A.}~\bibnamefont {Cavanna}}, \bibinfo {author} {\bibfnamefont {D.}~\bibnamefont {Mailly}},\ and\ \bibinfo {author} {\bibfnamefont {F.}~\bibnamefont {Pierre}},\ }\bibfield  {title} {\bibinfo {title} {Energy relaxation in the integer quantum hall regime},\ }\href {https://doi.org/10.1103/PhysRevLett.105.056803} {\bibfield  {journal} {\bibinfo  {journal} {Phys. Rev. Lett.}\ }\textbf {\bibinfo {volume} {105}},\ \bibinfo {pages} {056803} (\bibinfo {year} {2010})}\BibitemShut {NoStop}%
\bibitem [{\citenamefont {Koenig}\ \emph {et~al.}(2007)\citenamefont {Koenig}, \citenamefont {Wiedmann}, \citenamefont {Brune}, \citenamefont {Roth}, \citenamefont {Buhmann}, \citenamefont {Molenkamp}, \citenamefont {Qi},\ and\ \citenamefont {Zhang}}]{Konig2007SciSQH}%
  \BibitemOpen
  \bibfield  {author} {\bibinfo {author} {\bibfnamefont {M.}~\bibnamefont {Koenig}}, \bibinfo {author} {\bibfnamefont {S.}~\bibnamefont {Wiedmann}}, \bibinfo {author} {\bibfnamefont {C.}~\bibnamefont {Brune}}, \bibinfo {author} {\bibfnamefont {A.}~\bibnamefont {Roth}}, \bibinfo {author} {\bibfnamefont {H.}~\bibnamefont {Buhmann}}, \bibinfo {author} {\bibfnamefont {L.~W.}\ \bibnamefont {Molenkamp}}, \bibinfo {author} {\bibfnamefont {X.-L.}\ \bibnamefont {Qi}},\ and\ \bibinfo {author} {\bibfnamefont {S.-C.}\ \bibnamefont {Zhang}},\ }\bibfield  {title} {\bibinfo {title} {Quantum spin hall insulator state in hgte quantum wells},\ }\href {https://doi.org/10.1126/science.1148047} {\bibfield  {journal} {\bibinfo  {journal} {Science}\ }\textbf {\bibinfo {volume} {318}},\ \bibinfo {pages} {766} (\bibinfo {year} {2007})}\BibitemShut {NoStop}%
\bibitem [{\citenamefont {MacDonald}(1990)}]{McDo_fractional_edges_1990}%
  \BibitemOpen
  \bibfield  {author} {\bibinfo {author} {\bibfnamefont {A.~H.}\ \bibnamefont {MacDonald}},\ }\bibfield  {title} {\bibinfo {title} {Edge states in the fractional-quantum-hall-effect regime},\ }\href {https://doi.org/10.1103/PhysRevLett.64.220} {\bibfield  {journal} {\bibinfo  {journal} {Phys. Rev. Lett.}\ }\textbf {\bibinfo {volume} {64}},\ \bibinfo {pages} {220} (\bibinfo {year} {1990})}\BibitemShut {NoStop}%
\bibitem [{\citenamefont {Kane}\ \emph {et~al.}(1994)\citenamefont {Kane}, \citenamefont {Fisher},\ and\ \citenamefont {Polchinski}}]{KFP_equilibration_1994}%
  \BibitemOpen
  \bibfield  {author} {\bibinfo {author} {\bibfnamefont {C.~L.}\ \bibnamefont {Kane}}, \bibinfo {author} {\bibfnamefont {M.~P.~A.}\ \bibnamefont {Fisher}},\ and\ \bibinfo {author} {\bibfnamefont {J.}~\bibnamefont {Polchinski}},\ }\bibfield  {title} {\bibinfo {title} {Randomness at the edge: Theory of quantum hall transport at filling \ensuremath{\nu}=2/3},\ }\href {https://doi.org/10.1103/PhysRevLett.72.4129} {\bibfield  {journal} {\bibinfo  {journal} {Phys. Rev. Lett.}\ }\textbf {\bibinfo {volume} {72}},\ \bibinfo {pages} {4129} (\bibinfo {year} {1994})}\BibitemShut {NoStop}%
\bibitem [{\citenamefont {Protopopov}\ \emph {et~al.}(2017)\citenamefont {Protopopov}, \citenamefont {Gefen},\ and\ \citenamefont {Mirlin}}]{ProtopopovAnnals2017}%
  \BibitemOpen
  \bibfield  {author} {\bibinfo {author} {\bibfnamefont {I.}~\bibnamefont {Protopopov}}, \bibinfo {author} {\bibfnamefont {Y.}~\bibnamefont {Gefen}},\ and\ \bibinfo {author} {\bibfnamefont {A.}~\bibnamefont {Mirlin}},\ }\bibfield  {title} {\bibinfo {title} {Transport in a disordered $\nu$=2/3 fractional quantum hall junction},\ }\href {https://doi.org/https://doi.org/10.1016/j.aop.2017.07.015} {\bibfield  {journal} {\bibinfo  {journal} {Annals of Physics}\ }\textbf {\bibinfo {volume} {385}},\ \bibinfo {pages} {287} (\bibinfo {year} {2017})}\BibitemShut {NoStop}%
\bibitem [{\citenamefont {Nosiglia}\ \emph {et~al.}(2018)\citenamefont {Nosiglia}, \citenamefont {Park}, \citenamefont {Rosenow},\ and\ \citenamefont {Gefen}}]{NosigliaPRB2018}%
  \BibitemOpen
  \bibfield  {author} {\bibinfo {author} {\bibfnamefont {C.}~\bibnamefont {Nosiglia}}, \bibinfo {author} {\bibfnamefont {J.}~\bibnamefont {Park}}, \bibinfo {author} {\bibfnamefont {B.}~\bibnamefont {Rosenow}},\ and\ \bibinfo {author} {\bibfnamefont {Y.}~\bibnamefont {Gefen}},\ }\bibfield  {title} {\bibinfo {title} {Incoherent transport on the $\ensuremath{\nu}=2/3$ quantum hall edge},\ }\href {https://doi.org/10.1103/PhysRevB.98.115408} {\bibfield  {journal} {\bibinfo  {journal} {Phys. Rev. B}\ }\textbf {\bibinfo {volume} {98}},\ \bibinfo {pages} {115408} (\bibinfo {year} {2018})}\BibitemShut {NoStop}%
\bibitem [{\citenamefont {Park}\ \emph {et~al.}(2019)\citenamefont {Park}, \citenamefont {Mirlin}, \citenamefont {Rosenow},\ and\ \citenamefont {Gefen}}]{ParkPRB2019}%
  \BibitemOpen
  \bibfield  {author} {\bibinfo {author} {\bibfnamefont {J.}~\bibnamefont {Park}}, \bibinfo {author} {\bibfnamefont {A.~D.}\ \bibnamefont {Mirlin}}, \bibinfo {author} {\bibfnamefont {B.}~\bibnamefont {Rosenow}},\ and\ \bibinfo {author} {\bibfnamefont {Y.}~\bibnamefont {Gefen}},\ }\bibfield  {title} {\bibinfo {title} {Noise on complex quantum hall edges: Chiral anomaly and heat diffusion},\ }\href {https://doi.org/10.1103/PhysRevB.99.161302} {\bibfield  {journal} {\bibinfo  {journal} {Phys. Rev. B}\ }\textbf {\bibinfo {volume} {99}},\ \bibinfo {pages} {161302} (\bibinfo {year} {2019})}\BibitemShut {NoStop}%
\bibitem [{\citenamefont {Sp\aa{}nsl\"att}\ \emph {et~al.}(2019)\citenamefont {Sp\aa{}nsl\"att}, \citenamefont {Park}, \citenamefont {Gefen},\ and\ \citenamefont {Mirlin}}]{SpanslattPRL2019}%
  \BibitemOpen
  \bibfield  {author} {\bibinfo {author} {\bibfnamefont {C.}~\bibnamefont {Sp\aa{}nsl\"att}}, \bibinfo {author} {\bibfnamefont {J.}~\bibnamefont {Park}}, \bibinfo {author} {\bibfnamefont {Y.}~\bibnamefont {Gefen}},\ and\ \bibinfo {author} {\bibfnamefont {A.~D.}\ \bibnamefont {Mirlin}},\ }\bibfield  {title} {\bibinfo {title} {Topological classification of shot noise on fractional quantum hall edges},\ }\href {https://doi.org/10.1103/PhysRevLett.123.137701} {\bibfield  {journal} {\bibinfo  {journal} {Phys. Rev. Lett.}\ }\textbf {\bibinfo {volume} {123}},\ \bibinfo {pages} {137701} (\bibinfo {year} {2019})}\BibitemShut {NoStop}%
\bibitem [{\citenamefont {Sp\aa{}nsl\"att}\ \emph {et~al.}(2020)\citenamefont {Sp\aa{}nsl\"att}, \citenamefont {Park}, \citenamefont {Gefen},\ and\ \citenamefont {Mirlin}}]{SpanslattPRB2020}%
  \BibitemOpen
  \bibfield  {author} {\bibinfo {author} {\bibfnamefont {C.}~\bibnamefont {Sp\aa{}nsl\"att}}, \bibinfo {author} {\bibfnamefont {J.}~\bibnamefont {Park}}, \bibinfo {author} {\bibfnamefont {Y.}~\bibnamefont {Gefen}},\ and\ \bibinfo {author} {\bibfnamefont {A.~D.}\ \bibnamefont {Mirlin}},\ }\bibfield  {title} {\bibinfo {title} {Conductance plateaus and shot noise in fractional quantum hall point contacts},\ }\href {https://doi.org/10.1103/PhysRevB.101.075308} {\bibfield  {journal} {\bibinfo  {journal} {Phys. Rev. B}\ }\textbf {\bibinfo {volume} {101}},\ \bibinfo {pages} {075308} (\bibinfo {year} {2020})}\BibitemShut {NoStop}%
\bibitem [{\citenamefont {Fujisawa}\ and\ \citenamefont {Lin}(2021)}]{FujisawaPRB2021}%
  \BibitemOpen
  \bibfield  {author} {\bibinfo {author} {\bibfnamefont {T.}~\bibnamefont {Fujisawa}}\ and\ \bibinfo {author} {\bibfnamefont {C.}~\bibnamefont {Lin}},\ }\bibfield  {title} {\bibinfo {title} {Plasmon modes of coupled quantum hall edge channels in the presence of disorder-induced tunneling},\ }\href {https://doi.org/10.1103/PhysRevB.103.165302} {\bibfield  {journal} {\bibinfo  {journal} {Phys. Rev. B}\ }\textbf {\bibinfo {volume} {103}},\ \bibinfo {pages} {165302} (\bibinfo {year} {2021})}\BibitemShut {NoStop}%
\bibitem [{\citenamefont {Glattli}\ \emph {et~al.}(2024)\citenamefont {Glattli}, \citenamefont {Boudet}, \citenamefont {De},\ and\ \citenamefont {Roulleau}}]{Glattli2024arxiv}%
  \BibitemOpen
  \bibfield  {author} {\bibinfo {author} {\bibfnamefont {D.~C.}\ \bibnamefont {Glattli}}, \bibinfo {author} {\bibfnamefont {C.}~\bibnamefont {Boudet}}, \bibinfo {author} {\bibfnamefont {A.}~\bibnamefont {De}},\ and\ \bibinfo {author} {\bibfnamefont {P.}~\bibnamefont {Roulleau}},\ }\href {https://arxiv.org/abs/2407.07208} {\bibinfo {title} {Revisiting the physics of hole-conjugate fractional quantum hall channels}} (\bibinfo {year} {2024}),\ \Eprint {https://arxiv.org/abs/2407.07208} {arXiv:2407.07208 [cond-mat.mes-hall]} \BibitemShut {NoStop}%
\bibitem [{\citenamefont {Bid}\ \emph {et~al.}(2010)\citenamefont {Bid}, \citenamefont {Ofek}, \citenamefont {Inoue}, \citenamefont {Heiblum}, \citenamefont {Kane}, \citenamefont {Umansky},\ and\ \citenamefont {Mahalu}}]{Bid2010}%
  \BibitemOpen
  \bibfield  {author} {\bibinfo {author} {\bibfnamefont {A.}~\bibnamefont {Bid}}, \bibinfo {author} {\bibfnamefont {N.}~\bibnamefont {Ofek}}, \bibinfo {author} {\bibfnamefont {H.}~\bibnamefont {Inoue}}, \bibinfo {author} {\bibfnamefont {M.}~\bibnamefont {Heiblum}}, \bibinfo {author} {\bibfnamefont {C.~L.}\ \bibnamefont {Kane}}, \bibinfo {author} {\bibfnamefont {V.}~\bibnamefont {Umansky}},\ and\ \bibinfo {author} {\bibfnamefont {D.}~\bibnamefont {Mahalu}},\ }\bibfield  {title} {\bibinfo {title} {Observation of neutral modes in the fractional quantum hall regime},\ }\href {https://doi.org/10.1038/nature09277} {\bibfield  {journal} {\bibinfo  {journal} {Nature}\ }\textbf {\bibinfo {volume} {466}},\ \bibinfo {pages} {585} (\bibinfo {year} {2010})}\BibitemShut {NoStop}%
\bibitem [{\citenamefont {Lin}\ \emph {et~al.}(2019)\citenamefont {Lin}, \citenamefont {Eguchi}, \citenamefont {Hashisaka}, \citenamefont {Akiho}, \citenamefont {Muraki},\ and\ \citenamefont {Fujisawa}}]{Lin2019PRB}%
  \BibitemOpen
  \bibfield  {author} {\bibinfo {author} {\bibfnamefont {C.}~\bibnamefont {Lin}}, \bibinfo {author} {\bibfnamefont {R.}~\bibnamefont {Eguchi}}, \bibinfo {author} {\bibfnamefont {M.}~\bibnamefont {Hashisaka}}, \bibinfo {author} {\bibfnamefont {T.}~\bibnamefont {Akiho}}, \bibinfo {author} {\bibfnamefont {K.}~\bibnamefont {Muraki}},\ and\ \bibinfo {author} {\bibfnamefont {T.}~\bibnamefont {Fujisawa}},\ }\bibfield  {title} {\bibinfo {title} {Charge equilibration in integer and fractional quantum hall edge channels in a generalized hall-bar device},\ }\href {https://doi.org/10.1103/PhysRevB.99.195304} {\bibfield  {journal} {\bibinfo  {journal} {Phys. Rev. B}\ }\textbf {\bibinfo {volume} {99}},\ \bibinfo {pages} {195304} (\bibinfo {year} {2019})}\BibitemShut {NoStop}%
\bibitem [{\citenamefont {Cohen}\ \emph {et~al.}(2019)\citenamefont {Cohen}, \citenamefont {Ronen}, \citenamefont {Yang}, \citenamefont {Banitt}, \citenamefont {Park}, \citenamefont {Heiblum}, \citenamefont {Mirlin}, \citenamefont {Gefen},\ and\ \citenamefont {Umansky}}]{Cohen2019NComms}%
  \BibitemOpen
  \bibfield  {author} {\bibinfo {author} {\bibfnamefont {Y.}~\bibnamefont {Cohen}}, \bibinfo {author} {\bibfnamefont {Y.}~\bibnamefont {Ronen}}, \bibinfo {author} {\bibfnamefont {W.}~\bibnamefont {Yang}}, \bibinfo {author} {\bibfnamefont {D.}~\bibnamefont {Banitt}}, \bibinfo {author} {\bibfnamefont {J.}~\bibnamefont {Park}}, \bibinfo {author} {\bibfnamefont {M.}~\bibnamefont {Heiblum}}, \bibinfo {author} {\bibfnamefont {A.~D.}\ \bibnamefont {Mirlin}}, \bibinfo {author} {\bibfnamefont {Y.}~\bibnamefont {Gefen}},\ and\ \bibinfo {author} {\bibfnamefont {V.}~\bibnamefont {Umansky}},\ }\bibfield  {title} {\bibinfo {title} {Synthesizing a $\nu$=2/3 fractional quantum hall effect edge state from counter-propagating $\nu$=1 and $\nu$=1/3 states},\ }\href@noop {} {\bibfield  {journal} {\bibinfo  {journal} {Nature Communications}\ }\textbf {\bibinfo {volume} {10}},\ \bibinfo {pages} {1920} (\bibinfo {year} {2019})}\BibitemShut {NoStop}%
\bibitem [{\citenamefont {Lafont}\ \emph {et~al.}(2019)\citenamefont {Lafont}, \citenamefont {Rosenblatt}, \citenamefont {Heiblum},\ and\ \citenamefont {Umansky}}]{Lafont2019Science}%
  \BibitemOpen
  \bibfield  {author} {\bibinfo {author} {\bibfnamefont {F.}~\bibnamefont {Lafont}}, \bibinfo {author} {\bibfnamefont {A.}~\bibnamefont {Rosenblatt}}, \bibinfo {author} {\bibfnamefont {M.}~\bibnamefont {Heiblum}},\ and\ \bibinfo {author} {\bibfnamefont {V.}~\bibnamefont {Umansky}},\ }\bibfield  {title} {\bibinfo {title} {Counter-propagating charge transport in the quantum hall effect regime},\ }\href {https://doi.org/10.1126/science.aar3766} {\bibfield  {journal} {\bibinfo  {journal} {Science}\ }\textbf {\bibinfo {volume} {363}},\ \bibinfo {pages} {54} (\bibinfo {year} {2019})}\BibitemShut {NoStop}%
\bibitem [{\citenamefont {Srivastav}\ \emph {et~al.}(2021)\citenamefont {Srivastav}, \citenamefont {Kumar}, \citenamefont {Sp\aa{}nsl\"att}, \citenamefont {Watanabe}, \citenamefont {Taniguchi}, \citenamefont {Mirlin}, \citenamefont {Gefen},\ and\ \citenamefont {Das}}]{SrivastavPRL2021}%
  \BibitemOpen
  \bibfield  {author} {\bibinfo {author} {\bibfnamefont {S.~K.}\ \bibnamefont {Srivastav}}, \bibinfo {author} {\bibfnamefont {R.}~\bibnamefont {Kumar}}, \bibinfo {author} {\bibfnamefont {C.}~\bibnamefont {Sp\aa{}nsl\"att}}, \bibinfo {author} {\bibfnamefont {K.}~\bibnamefont {Watanabe}}, \bibinfo {author} {\bibfnamefont {T.}~\bibnamefont {Taniguchi}}, \bibinfo {author} {\bibfnamefont {A.~D.}\ \bibnamefont {Mirlin}}, \bibinfo {author} {\bibfnamefont {Y.}~\bibnamefont {Gefen}},\ and\ \bibinfo {author} {\bibfnamefont {A.}~\bibnamefont {Das}},\ }\bibfield  {title} {\bibinfo {title} {Vanishing thermal equilibration for hole-conjugate fractional quantum hall states in graphene},\ }\href {https://doi.org/10.1103/PhysRevLett.126.216803} {\bibfield  {journal} {\bibinfo  {journal} {Phys. Rev. Lett.}\ }\textbf {\bibinfo {volume} {126}},\ \bibinfo {pages} {216803} (\bibinfo {year} {2021})}\BibitemShut {NoStop}%
\bibitem [{\citenamefont {Melcer}\ \emph {et~al.}(2022)\citenamefont {Melcer}, \citenamefont {Dutta}, \citenamefont {Sp{\aa}nsl{\"a}tt}, \citenamefont {Park}, \citenamefont {Mirlin},\ and\ \citenamefont {Umansky}}]{Melcer2022NatComms}%
  \BibitemOpen
  \bibfield  {author} {\bibinfo {author} {\bibfnamefont {R.~A.}\ \bibnamefont {Melcer}}, \bibinfo {author} {\bibfnamefont {B.}~\bibnamefont {Dutta}}, \bibinfo {author} {\bibfnamefont {C.}~\bibnamefont {Sp{\aa}nsl{\"a}tt}}, \bibinfo {author} {\bibfnamefont {J.}~\bibnamefont {Park}}, \bibinfo {author} {\bibfnamefont {A.~D.}\ \bibnamefont {Mirlin}},\ and\ \bibinfo {author} {\bibfnamefont {V.}~\bibnamefont {Umansky}},\ }\bibfield  {title} {\bibinfo {title} {Absent thermal equilibration on fractional quantum hall edges over macroscopic scale},\ }\href {https://doi.org/10.1038/s41467-022-28009-0} {\bibfield  {journal} {\bibinfo  {journal} {Nature Communications}\ }\textbf {\bibinfo {volume} {13}},\ \bibinfo {pages} {376} (\bibinfo {year} {2022})}\BibitemShut {NoStop}%
\bibitem [{\citenamefont {Roth}\ \emph {et~al.}(2009)\citenamefont {Roth}, \citenamefont {Brune}, \citenamefont {Buhmann}, \citenamefont {Molenkamp}, \citenamefont {Maciejko}, \citenamefont {Qi},\ and\ \citenamefont {Zhang}}]{Roth2009Science}%
  \BibitemOpen
  \bibfield  {author} {\bibinfo {author} {\bibfnamefont {A.}~\bibnamefont {Roth}}, \bibinfo {author} {\bibfnamefont {C.}~\bibnamefont {Brune}}, \bibinfo {author} {\bibfnamefont {H.}~\bibnamefont {Buhmann}}, \bibinfo {author} {\bibfnamefont {L.~W.}\ \bibnamefont {Molenkamp}}, \bibinfo {author} {\bibfnamefont {J.}~\bibnamefont {Maciejko}}, \bibinfo {author} {\bibfnamefont {X.-L.}\ \bibnamefont {Qi}},\ and\ \bibinfo {author} {\bibfnamefont {S.-C.}\ \bibnamefont {Zhang}},\ }\bibfield  {title} {\bibinfo {title} {Nonlocal transport in the quantum spin hall state},\ }\href {https://doi.org/10.1126/science.1174736} {\bibfield  {journal} {\bibinfo  {journal} {Science}\ }\textbf {\bibinfo {volume} {325}},\ \bibinfo {pages} {294} (\bibinfo {year} {2009})}\BibitemShut {NoStop}%
\bibitem [{\citenamefont {Knez}\ \emph {et~al.}(2011)\citenamefont {Knez}, \citenamefont {Du},\ and\ \citenamefont {Sullivan}}]{Knez2011}%
  \BibitemOpen
  \bibfield  {author} {\bibinfo {author} {\bibfnamefont {I.}~\bibnamefont {Knez}}, \bibinfo {author} {\bibfnamefont {R.-R.}\ \bibnamefont {Du}},\ and\ \bibinfo {author} {\bibfnamefont {G.}~\bibnamefont {Sullivan}},\ }\bibfield  {title} {\bibinfo {title} {{Evidence for Helical Edge Modes in Inverted $\mathrm{InAs}/\mathrm{GaSb}$ Quantum Wells}},\ }\href {https://doi.org/10.1103/PhysRevLett.107.136603} {\bibfield  {journal} {\bibinfo  {journal} {Phys. Rev. Lett.}\ }\textbf {\bibinfo {volume} {107}},\ \bibinfo {pages} {136603} (\bibinfo {year} {2011})}\BibitemShut {NoStop}%
\bibitem [{\citenamefont {Wu}\ \emph {et~al.}(2018)\citenamefont {Wu}, \citenamefont {Fatemi}, \citenamefont {Gibson}, \citenamefont {Watanabe}, \citenamefont {Taniguchi}, \citenamefont {Cava},\ and\ \citenamefont {Jarillo-Herrero}}]{Wu2018QSH}%
  \BibitemOpen
  \bibfield  {author} {\bibinfo {author} {\bibfnamefont {S.}~\bibnamefont {Wu}}, \bibinfo {author} {\bibfnamefont {V.}~\bibnamefont {Fatemi}}, \bibinfo {author} {\bibfnamefont {Q.~D.}\ \bibnamefont {Gibson}}, \bibinfo {author} {\bibfnamefont {K.}~\bibnamefont {Watanabe}}, \bibinfo {author} {\bibfnamefont {T.}~\bibnamefont {Taniguchi}}, \bibinfo {author} {\bibfnamefont {R.~J.}\ \bibnamefont {Cava}},\ and\ \bibinfo {author} {\bibfnamefont {P.}~\bibnamefont {Jarillo-Herrero}},\ }\bibfield  {title} {\bibinfo {title} {Observation of the quantum spin hall effect up to 100 kelvin in a monolayer crystal},\ }\href {https://doi.org/10.1126/science.aan6003} {\bibfield  {journal} {\bibinfo  {journal} {Science}\ }\textbf {\bibinfo {volume} {359}},\ \bibinfo {pages} {76} (\bibinfo {year} {2018})}\BibitemShut {NoStop}%
\bibitem [{\citenamefont {Veyrat}\ \emph {et~al.}(2020)\citenamefont {Veyrat}, \citenamefont {Déprez}, \citenamefont {Coissard}, \citenamefont {Li}, \citenamefont {Gay}, \citenamefont {Watanabe}, \citenamefont {Taniguchi}, \citenamefont {Han}, \citenamefont {Piot}, \citenamefont {Sellier},\ and\ \citenamefont {Sacépé}}]{Veyrat2020Science}%
  \BibitemOpen
  \bibfield  {author} {\bibinfo {author} {\bibfnamefont {L.}~\bibnamefont {Veyrat}}, \bibinfo {author} {\bibfnamefont {C.}~\bibnamefont {Déprez}}, \bibinfo {author} {\bibfnamefont {A.}~\bibnamefont {Coissard}}, \bibinfo {author} {\bibfnamefont {X.}~\bibnamefont {Li}}, \bibinfo {author} {\bibfnamefont {F.}~\bibnamefont {Gay}}, \bibinfo {author} {\bibfnamefont {K.}~\bibnamefont {Watanabe}}, \bibinfo {author} {\bibfnamefont {T.}~\bibnamefont {Taniguchi}}, \bibinfo {author} {\bibfnamefont {Z.}~\bibnamefont {Han}}, \bibinfo {author} {\bibfnamefont {B.~A.}\ \bibnamefont {Piot}}, \bibinfo {author} {\bibfnamefont {H.}~\bibnamefont {Sellier}},\ and\ \bibinfo {author} {\bibfnamefont {B.}~\bibnamefont {Sacépé}},\ }\bibfield  {title} {\bibinfo {title} {Helical quantum hall phase in graphene on $\mathrm{SrTiO_3}$},\ }\href {https://doi.org/10.1126/science.aax8201} {\bibfield  {journal} {\bibinfo  {journal} {Science}\ }\textbf {\bibinfo {volume} {367}},\ \bibinfo {pages} {781} (\bibinfo {year} {2020})}\BibitemShut
  {NoStop}%
\bibitem [{\citenamefont {Kang}\ \emph {et~al.}(2024)\citenamefont {Kang}, \citenamefont {Shen}, \citenamefont {Qiu}, \citenamefont {Zeng}, \citenamefont {Xia}, \citenamefont {Watanabe}, \citenamefont {Taniguchi}, \citenamefont {Shan},\ and\ \citenamefont {Mak}}]{Kang2024}%
  \BibitemOpen
  \bibfield  {author} {\bibinfo {author} {\bibfnamefont {K.}~\bibnamefont {Kang}}, \bibinfo {author} {\bibfnamefont {B.}~\bibnamefont {Shen}}, \bibinfo {author} {\bibfnamefont {Y.}~\bibnamefont {Qiu}}, \bibinfo {author} {\bibfnamefont {Y.}~\bibnamefont {Zeng}}, \bibinfo {author} {\bibfnamefont {Z.}~\bibnamefont {Xia}}, \bibinfo {author} {\bibfnamefont {K.}~\bibnamefont {Watanabe}}, \bibinfo {author} {\bibfnamefont {T.}~\bibnamefont {Taniguchi}}, \bibinfo {author} {\bibfnamefont {J.}~\bibnamefont {Shan}},\ and\ \bibinfo {author} {\bibfnamefont {K.~F.}\ \bibnamefont {Mak}},\ }\bibfield  {title} {\bibinfo {title} {{Evidence of the fractional quantum spin Hall effect in moir{\ifmmode\acute{e}\else\'{e}\fi} MoTe2}},\ }\href {https://doi.org/10.1038/s41586-024-07214-5} {\bibfield  {journal} {\bibinfo  {journal} {Nature}\ }\textbf {\bibinfo {volume} {628}},\ \bibinfo {pages} {522} (\bibinfo {year} {2024})}\BibitemShut {NoStop}%
\bibitem [{\citenamefont {Yang}\ \emph {et~al.}(2024)\citenamefont {Yang}, \citenamefont {Bhujel}, \citenamefont {Chica}, \citenamefont {Telford}, \citenamefont {Roy}, \citenamefont {Ibrahim}, \citenamefont {Chshiev}, \citenamefont {Cosset-Ch{\'e}neau},\ and\ \citenamefont {Wees}}]{Yang2024NComms}%
  \BibitemOpen
  \bibfield  {author} {\bibinfo {author} {\bibfnamefont {B.}~\bibnamefont {Yang}}, \bibinfo {author} {\bibfnamefont {B.}~\bibnamefont {Bhujel}}, \bibinfo {author} {\bibfnamefont {D.~G.}\ \bibnamefont {Chica}}, \bibinfo {author} {\bibfnamefont {E.~J.}\ \bibnamefont {Telford}}, \bibinfo {author} {\bibfnamefont {X.}~\bibnamefont {Roy}}, \bibinfo {author} {\bibfnamefont {F.}~\bibnamefont {Ibrahim}}, \bibinfo {author} {\bibfnamefont {M.}~\bibnamefont {Chshiev}}, \bibinfo {author} {\bibfnamefont {M.}~\bibnamefont {Cosset-Ch{\'e}neau}},\ and\ \bibinfo {author} {\bibfnamefont {B.~J.~v.}\ \bibnamefont {Wees}},\ }\bibfield  {title} {\bibinfo {title} {Electrostatically controlled spin polarization in graphene-crsbr magnetic proximity heterostructures},\ }\href {https://doi.org/10.1038/s41467-024-48809-w} {\bibfield  {journal} {\bibinfo  {journal} {Nature Communications}\ }\textbf {\bibinfo {volume} {15}},\ \bibinfo {pages} {4459} (\bibinfo {year} {2024})}\BibitemShut {NoStop}%
\bibitem [{\citenamefont {Banerjee}\ \emph {et~al.}(2017)\citenamefont {Banerjee}, \citenamefont {Heiblum}, \citenamefont {Rosenblatt}, \citenamefont {Oreg}, \citenamefont {Feldman}, \citenamefont {Stern},\ and\ \citenamefont {Umansky}}]{Banerjee2017Nature}%
  \BibitemOpen
  \bibfield  {author} {\bibinfo {author} {\bibfnamefont {M.}~\bibnamefont {Banerjee}}, \bibinfo {author} {\bibfnamefont {M.}~\bibnamefont {Heiblum}}, \bibinfo {author} {\bibfnamefont {A.}~\bibnamefont {Rosenblatt}}, \bibinfo {author} {\bibfnamefont {Y.}~\bibnamefont {Oreg}}, \bibinfo {author} {\bibfnamefont {D.~E.}\ \bibnamefont {Feldman}}, \bibinfo {author} {\bibfnamefont {A.}~\bibnamefont {Stern}},\ and\ \bibinfo {author} {\bibfnamefont {V.}~\bibnamefont {Umansky}},\ }\bibfield  {title} {\bibinfo {title} {Observed quantization of anyonic heat flow},\ }\href {https://doi.org/10.1038/nature22052} {\bibfield  {journal} {\bibinfo  {journal} {Nature}\ }\textbf {\bibinfo {volume} {545}},\ \bibinfo {pages} {75} (\bibinfo {year} {2017})}\BibitemShut {NoStop}%
\bibitem [{\citenamefont {Dean}\ \emph {et~al.}(2010)\citenamefont {Dean}, \citenamefont {Young}, \citenamefont {Meric}, \citenamefont {Lee}, \citenamefont {Wang}, \citenamefont {Sorgenfrei}, \citenamefont {Watanabe}, \citenamefont {Taniguchi}, \citenamefont {Kim}, \citenamefont {Shepard},\ and\ \citenamefont {Hone}}]{Dean2010NatNano}%
  \BibitemOpen
  \bibfield  {author} {\bibinfo {author} {\bibfnamefont {C.~R.}\ \bibnamefont {Dean}}, \bibinfo {author} {\bibfnamefont {A.~F.}\ \bibnamefont {Young}}, \bibinfo {author} {\bibfnamefont {I.}~\bibnamefont {Meric}}, \bibinfo {author} {\bibfnamefont {C.}~\bibnamefont {Lee}}, \bibinfo {author} {\bibfnamefont {L.}~\bibnamefont {Wang}}, \bibinfo {author} {\bibfnamefont {S.}~\bibnamefont {Sorgenfrei}}, \bibinfo {author} {\bibfnamefont {K.}~\bibnamefont {Watanabe}}, \bibinfo {author} {\bibfnamefont {T.}~\bibnamefont {Taniguchi}}, \bibinfo {author} {\bibfnamefont {P.}~\bibnamefont {Kim}}, \bibinfo {author} {\bibfnamefont {K.~L.}\ \bibnamefont {Shepard}},\ and\ \bibinfo {author} {\bibfnamefont {J.}~\bibnamefont {Hone}},\ }\bibfield  {title} {\bibinfo {title} {Boron nitride substrates for high-quality graphene electronics},\ }\href {https://doi.org/10.1038/nnano.2010.172} {\bibfield  {journal} {\bibinfo  {journal} {Nature Nanotechnology}\ }\textbf {\bibinfo {volume} {5}},\ \bibinfo {pages} {722} (\bibinfo {year}
  {2010})}\BibitemShut {NoStop}%
\bibitem [{SM()}]{SM}%
  \BibitemOpen
  \href@noop {} {\bibinfo {title} {See supplemental materials}}\BibitemShut {NoStop}%
\bibitem [{\citenamefont {Kane}\ and\ \citenamefont {Fisher}(1997)}]{KaneFisherPRB95}%
  \BibitemOpen
  \bibfield  {author} {\bibinfo {author} {\bibfnamefont {C.~L.}\ \bibnamefont {Kane}}\ and\ \bibinfo {author} {\bibfnamefont {M.~P.~A.}\ \bibnamefont {Fisher}},\ }\bibfield  {title} {\bibinfo {title} {Quantized thermal transport in the fractional quantum hall effect},\ }\href {https://doi.org/10.1103/PhysRevB.55.15832} {\bibfield  {journal} {\bibinfo  {journal} {Phys. Rev. B}\ }\textbf {\bibinfo {volume} {55}},\ \bibinfo {pages} {15832} (\bibinfo {year} {1997})}\BibitemShut {NoStop}%
\bibitem [{\citenamefont {Hashisaka}\ \emph {et~al.}(2023)\citenamefont {Hashisaka}, \citenamefont {Ito}, \citenamefont {Akiho}, \citenamefont {Sasaki}, \citenamefont {Kumada}, \citenamefont {Shibata},\ and\ \citenamefont {Muraki}}]{HashisakaPRX2023}%
  \BibitemOpen
  \bibfield  {author} {\bibinfo {author} {\bibfnamefont {M.}~\bibnamefont {Hashisaka}}, \bibinfo {author} {\bibfnamefont {T.}~\bibnamefont {Ito}}, \bibinfo {author} {\bibfnamefont {T.}~\bibnamefont {Akiho}}, \bibinfo {author} {\bibfnamefont {S.}~\bibnamefont {Sasaki}}, \bibinfo {author} {\bibfnamefont {N.}~\bibnamefont {Kumada}}, \bibinfo {author} {\bibfnamefont {N.}~\bibnamefont {Shibata}},\ and\ \bibinfo {author} {\bibfnamefont {K.}~\bibnamefont {Muraki}},\ }\bibfield  {title} {\bibinfo {title} {Coherent-incoherent crossover of charge and neutral mode transport as evidence for the disorder-dominated fractional edge phase},\ }\href {https://doi.org/10.1103/PhysRevX.13.031024} {\bibfield  {journal} {\bibinfo  {journal} {Phys. Rev. X}\ }\textbf {\bibinfo {volume} {13}},\ \bibinfo {pages} {031024} (\bibinfo {year} {2023})}\BibitemShut {NoStop}%
\bibitem [{\citenamefont {Krizman}\ \emph {et~al.}(2024)\citenamefont {Krizman}, \citenamefont {Bermejo-Ortiz}, \citenamefont {Zakusylo}, \citenamefont {Hajlaoui}, \citenamefont {Takashiro}, \citenamefont {Rosmus}, \citenamefont {Olszowska}, \citenamefont {Ko\l{}odziej}, \citenamefont {Bauer}, \citenamefont {Guldner}, \citenamefont {Springholz},\ and\ \citenamefont {de~Vaulchier}}]{KrizmanPRL2024}%
  \BibitemOpen
  \bibfield  {author} {\bibinfo {author} {\bibfnamefont {G.}~\bibnamefont {Krizman}}, \bibinfo {author} {\bibfnamefont {J.}~\bibnamefont {Bermejo-Ortiz}}, \bibinfo {author} {\bibfnamefont {T.}~\bibnamefont {Zakusylo}}, \bibinfo {author} {\bibfnamefont {M.}~\bibnamefont {Hajlaoui}}, \bibinfo {author} {\bibfnamefont {T.}~\bibnamefont {Takashiro}}, \bibinfo {author} {\bibfnamefont {M.}~\bibnamefont {Rosmus}}, \bibinfo {author} {\bibfnamefont {N.}~\bibnamefont {Olszowska}}, \bibinfo {author} {\bibfnamefont {J.~J.}\ \bibnamefont {Ko\l{}odziej}}, \bibinfo {author} {\bibfnamefont {G.}~\bibnamefont {Bauer}}, \bibinfo {author} {\bibfnamefont {Y.}~\bibnamefont {Guldner}}, \bibinfo {author} {\bibfnamefont {G.}~\bibnamefont {Springholz}},\ and\ \bibinfo {author} {\bibfnamefont {L.-A.}\ \bibnamefont {de~Vaulchier}},\ }\bibfield  {title} {\bibinfo {title} {Valley-polarized quantum hall phase in a strain-controlled dirac system},\ }\href {https://doi.org/10.1103/PhysRevLett.132.166601} {\bibfield  {journal} {\bibinfo
  {journal} {Phys. Rev. Lett.}\ }\textbf {\bibinfo {volume} {132}},\ \bibinfo {pages} {166601} (\bibinfo {year} {2024})}\BibitemShut {NoStop}%
\bibitem [{\citenamefont {Sivre}\ \emph {et~al.}(2018)\citenamefont {Sivre}, \citenamefont {Anthore}, \citenamefont {Parmentier}, \citenamefont {Cavanna}, \citenamefont {Gennser}, \citenamefont {Ouerghi}, \citenamefont {Jin},\ and\ \citenamefont {Pierre}}]{Sivre2018}%
  \BibitemOpen
  \bibfield  {author} {\bibinfo {author} {\bibfnamefont {E.}~\bibnamefont {Sivre}}, \bibinfo {author} {\bibfnamefont {A.}~\bibnamefont {Anthore}}, \bibinfo {author} {\bibfnamefont {F.~D.}\ \bibnamefont {Parmentier}}, \bibinfo {author} {\bibfnamefont {A.}~\bibnamefont {Cavanna}}, \bibinfo {author} {\bibfnamefont {U.}~\bibnamefont {Gennser}}, \bibinfo {author} {\bibfnamefont {A.}~\bibnamefont {Ouerghi}}, \bibinfo {author} {\bibfnamefont {Y.}~\bibnamefont {Jin}},\ and\ \bibinfo {author} {\bibfnamefont {F.}~\bibnamefont {Pierre}},\ }\bibfield  {title} {\bibinfo {title} {Heat coulomb blockade of one ballistic channel},\ }\href {https://doi.org/10.1038/nphys4280} {\bibfield  {journal} {\bibinfo  {journal} {Nature Physics}\ }\textbf {\bibinfo {volume} {14}},\ \bibinfo {pages} {145} (\bibinfo {year} {2018})}\BibitemShut {NoStop}%
\bibitem [{\citenamefont {St\"abler}\ and\ \citenamefont {Sukhorukov}(2023)}]{Stabler2023PRB}%
  \BibitemOpen
  \bibfield  {author} {\bibinfo {author} {\bibfnamefont {F.}~\bibnamefont {St\"abler}}\ and\ \bibinfo {author} {\bibfnamefont {E.}~\bibnamefont {Sukhorukov}},\ }\bibfield  {title} {\bibinfo {title} {Mesoscopic heat multiplier and fractionalizer},\ }\href {https://doi.org/10.1103/PhysRevB.108.235405} {\bibfield  {journal} {\bibinfo  {journal} {Phys. Rev. B}\ }\textbf {\bibinfo {volume} {108}},\ \bibinfo {pages} {235405} (\bibinfo {year} {2023})}\BibitemShut {NoStop}%
\bibitem [{\citenamefont {Staebler}\ \emph {et~al.}(2024)\citenamefont {Staebler}, \citenamefont {Gadiaga},\ and\ \citenamefont {Sukhorukov}}]{stabler2024giantheatfluxeffect}%
  \BibitemOpen
  \bibfield  {author} {\bibinfo {author} {\bibfnamefont {F.}~\bibnamefont {Staebler}}, \bibinfo {author} {\bibfnamefont {A.}~\bibnamefont {Gadiaga}},\ and\ \bibinfo {author} {\bibfnamefont {E.~V.}\ \bibnamefont {Sukhorukov}},\ }\href {https://arxiv.org/abs/2411.11495} {\bibinfo {title} {Giant heat flux effect in non-chiral transmission lines}} (\bibinfo {year} {2024}),\ \Eprint {https://arxiv.org/abs/2411.11495} {arXiv:2411.11495 [cond-mat.mes-hall]} \BibitemShut {NoStop}%
\bibitem [{\citenamefont {Hashisaka}\ \emph {et~al.}(2021)\citenamefont {Hashisaka}, \citenamefont {Jonckheere}, \citenamefont {Akiho}, \citenamefont {Sasaki}, \citenamefont {Rech}, \citenamefont {Martin},\ and\ \citenamefont {Muraki}}]{Hashisaka2021NComms}%
  \BibitemOpen
  \bibfield  {author} {\bibinfo {author} {\bibfnamefont {M.}~\bibnamefont {Hashisaka}}, \bibinfo {author} {\bibfnamefont {T.}~\bibnamefont {Jonckheere}}, \bibinfo {author} {\bibfnamefont {T.}~\bibnamefont {Akiho}}, \bibinfo {author} {\bibfnamefont {S.}~\bibnamefont {Sasaki}}, \bibinfo {author} {\bibfnamefont {J.}~\bibnamefont {Rech}}, \bibinfo {author} {\bibfnamefont {T.}~\bibnamefont {Martin}},\ and\ \bibinfo {author} {\bibfnamefont {K.}~\bibnamefont {Muraki}},\ }\bibfield  {title} {\bibinfo {title} {Andreev reflection of fractional quantum hall quasiparticles},\ }\href {https://doi.org/10.1038/s41467-021-23160-6} {\bibfield  {journal} {\bibinfo  {journal} {Nature Communications}\ }\textbf {\bibinfo {volume} {12}},\ \bibinfo {pages} {2794} (\bibinfo {year} {2021})}\BibitemShut {NoStop}%
\end{thebibliography}%

\newpage
\onecolumngrid
\setcounter{figure}{0}
\renewcommand{\thepage}{S\arabic{page}} 
\renewcommand{\thesection}{S\arabic{section}}  
\renewcommand{\thetable}{S\arabic{table}}  
\renewcommand{\thefigure}{S\arabic{figure}} 
\renewcommand{\theequation}{S.\arabic{equation}}
\title{Supplementary Material for ``Ballistic-to-diffusive transition in engineered counter-propagating quantum Hall channels"}
\maketitle
\section{Sample description}

\par
\begin{figure}[hbt!]
    \centering
    \includegraphics[width=1.0\linewidth]{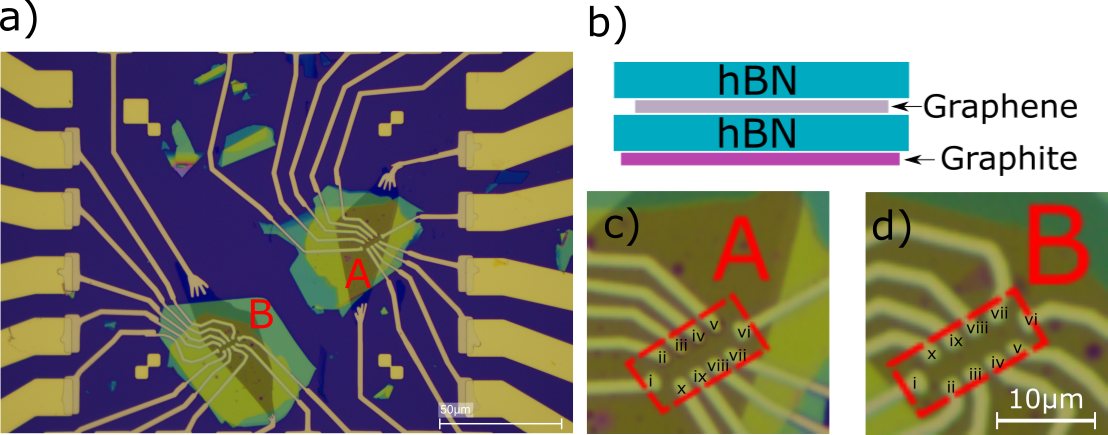}
    \caption{a) Optical microscope picture of Hall bars A and B. b) Side view schematic of each Hall bar. c,d) Zoomed optical pictures of Hall bars A and B. We connect the contacts of the same number together to form an effective Hall bar with counter-propagating channels. }
    \label{fig:sample}
\end{figure}

 Our sample (see Fig. \ref{fig:sample}) consists of two individually graphite-gated Hall bars (A and B) made of hexagonal Boron Nitride (hBN) encapsulated monolayer graphene, deposited on the same $SiO_2$ chip pre-patterned with Au/Cr pads. We then connect the sample to the pads by reactive ion etching (RIE), electron-beam lithography, and metal evaporation (Au/Pd/Cr). Each Hall bar has $10$ contacts numbered between $\mathrm{i}$ and $\mathrm{x}$. We then connect each pair of contacts using bonding wires, in such a way that the $k-$th contact of device A, starting clockwise from contact $\mathrm{i}$, is connected with the $k -$th contact of device B, starting anticlockwise from the contact $\mathrm{i}$. This will form an effective Hall bar with counter-propagating edge channels once placed under a perpendicular magnetic field. Note that in Hall bar A, contacts $\mathrm{ii}$ and $\mathrm{iii}$ are shorted together probably due to a piece of graphite on the substrate or by touching bonding wires, making 3 effective Ohmic contacts on the ``top" edge and 4 in the ``bottom". 
\par The following table shows the correspondence of contacts in each configuration in main text.
\begin{center}
\begin{tabular}{|c|c|c|c|c|c|c|c|c|c|}
\hline
 & i & ii/iii & iv & v & vi & vii & viii & ix & x \\
\hline
Asymmetric & 4 & 5 & 6 & 7 & 8 & 0 & 1 & 2 & 3 \\
\hline
Split & 0 & 1,t & 2,t & 3,t & 4,t/5,b & 4,b & 3,b & 2,b & 1,b \\
\hline
\end{tabular}
\end{center}

\section{Derivation of voltage profile and two-point conductance}

\par
\begin{figure}[hbt!]
    \centering
    \includegraphics[width=1.0\linewidth]{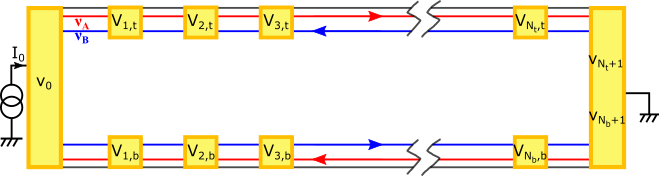}
    \caption{Schematic of our effective Hall bar with counter-propagating edge channels of filling factors $\nu_\mathrm{A}$ and $\nu_\mathrm{B}$.}
    \label{fig:circuit}
\end{figure}

 Here we express the voltages $V_{i,\gamma}$ on each contact $i$ in a Hall bar hosting counter-propagating edge channels each carrying 1 quantum of conductance $e^2/h$ (this can be extended easily to fractional values). $\nu_\mathrm{A}$ channels leave the source on the top edge, with $\nu_\mathrm{B}$ counter-propagating channels, as represented by Fig.\ref{fig:circuit}, with $\gamma =$ t,b (stand for ``top" and ``bottom" respectively). The top (bottom) edge contains $N_\mathrm{t}$($N_\mathrm{b}$) intermediate floating Ohmic contacts. We first write $V_{i,\gamma}$ as a function of $V_0$ and derive the two-point conductance $G_\mathrm{2w} = I_0/V_0$ which depends on $\nu_\mathrm{A}$,  $\nu_\mathrm{B}$ and number of pairs of contacts $N$ between source and drain (assuming $N_\mathrm{t}=N_\mathrm{b}=N$. We assume so here, because experimentally in the ``conductance" configuration, we always look at the symmetrical case by connecting the extra contacts to ground (see main text).

First we write current conservation at contact $V_{i,\mathrm{t}}$, assuming that the current emitted from the contact in one channel obeys Landauer-Büttiker formula $I_{\mathrm{1ch}}=G_QV_i$, with $G_Q=e^2/h$ the conductance quantum: 
\begin{equation*}
    V_{i,\mathrm{t}}(\nu_\mathrm{A}+\nu_\mathrm{B}) = \nu_\mathrm{A} V_{i-1,\mathrm{t}} + \nu_\mathrm{B} V_{i+1,\mathrm{t}}.
    \label{eq_LB}
\end{equation*}

We then define $\Delta V_{i,\gamma}=V_{i,\gamma}-V_{i-1,\gamma}$, which obeys the recurrence relation:
\begin{align}
    \Delta V_{i+1,\mathrm{t}}&=\frac{\nu_\mathrm{A}}{\nu_\mathrm{B}} \Delta V_{i,\mathrm{t}} = \left(\frac{\nu_\mathrm{A}}{\nu_\mathrm{B}} \right)^{i} \Delta V_{1,\mathrm{t}},
    \label{eq_deltaVtop}
\end{align}

and similarly for $\Delta V_{i,\mathrm{b}}$ up to a swap between $\nu_\mathrm{A}$ and $\nu_\mathrm{B}$. 
Knowing that:
\begin{align}
    V_{i,\gamma} - V_0  = \sum_{k=1}^{i} \Delta V_{k,\gamma} \label{eq_trick2} 
\end{align}

\par Using Eq. \ref{eq_deltaVtop} and that $V_{N_\mathrm{t}+1,\mathrm{t}}=0$, with geometric progression we have:

\begin{align}
\Delta V_{1,\mathrm{t}} &= \frac{\frac{\nu_\mathrm{A}}{\nu_\mathrm{B}}-1} {1-\left( \frac{\nu_\mathrm{A}}{\nu_\mathrm{B}}\right)^{N_\mathrm{t}+1}} V_0.
\end{align}
The same calculation applies to $\Delta V_{1,\mathrm{b}}$, with interchanged $\nu_\mathrm{A}$ and $\nu_\mathrm{B}$. This applies to $V_{i,\gamma}$ as well in the following text.


In the end, we obtain the voltage at each Landauer reservoir:
\begin{align}
V_{i,\mathrm{t}} &= \frac{\left( \frac{\nu_\mathrm{A}}{\nu_\mathrm{B}} \right)^i - \left( \frac{\nu_\mathrm{A}}{\nu_\mathrm{B}}\right)^{N_\mathrm{t}+1}}{1-\left( \frac{\nu_\mathrm{A}}{\nu_\mathrm{B}}\right)^{N_\mathrm{t}+1}} V_0.   \label{eq_Vi_top} 
\end{align}

\par As observed in the main text, $V_{i,\gamma}$ varies exponentially with contact index $i$.  Indeed, Eq. \ref{eq_Vi_top} can be rewritten as:
\begin{align}
    V_{i,\mathrm{t}} = \frac{1-\exp \left[ -\left(N_\mathrm{t}+1-i\right) \log \left(\frac{\nu_\mathrm{A}}{\nu_\mathrm{B}}\right) \right]}{1-\exp \left[ -(N_\mathrm{t}+1) \log\left( \frac{\nu_\mathrm{A}}{\nu_\mathrm{B}} \right)\right]} V_0. \label{eq_Vi_top_exp} \\ \nonumber
\end{align}

\par When $\nu_\mathrm{A}=\nu_\mathrm{B}$, the dimensionless equilibration length $\delta=1/|\log(\nu_\mathrm{A}/\nu_\mathrm{B})|$ diverges. 
The singular case $\nu_\mathrm{A}=\nu_\mathrm{B}$ can be treated by considering $\nu_\mathrm{A}=\nu_\mathrm{B}+\epsilon$ with $\epsilon \ll 1$. Expanding at first order, then taking $\epsilon=0$, we get:

\begin{align}
    V_{i,\gamma} = \left( 1-\frac{i}{N_\gamma+1}\right) V_0, \label{eq_Vi_top_equals_bottom}
\end{align}

that is, an Ohmic profile. This relation can also be easily obtained by writing again current conservation with $\nu_\mathrm{A}=\nu_\mathrm{B}$ from the beginning. 

\par We can finally calculate two point conductance $G_\mathrm{2w} = I_0/V_0$ using current conservation on contact 0, it depends purely on filling factors $\nu_{\mathrm{A,B}}$ and the number of pairs of intermediate contacts $N$.
\begin{align}
    &I_0 +G_Q(\nu_\mathrm{B}V_{1,\mathrm{t}}+\nu_\mathrm{A}V_{1,\mathrm{b}}) = G_Q V_0 (\nu_\mathrm{A}+\nu_\mathrm{B}) \nonumber \\
    &G_\mathrm{2w} = \frac{I_0}{V_0} = G_Q (\nu_\mathrm{A}-\nu_\mathrm{B})\frac{1+\left(\frac{\nu_\mathrm{B}}{\nu_\mathrm{A}}\right)^{N+1}}{1-\left(\frac{\nu_\mathrm{B}}{\nu_\mathrm{A}}\right)^{N+1}} \label{eq_G}
\end{align}


\par The two-point conductance can also be written in the following way, with $G_\infty = G_Q|\nu_\mathrm{A}-\nu_\mathrm{B}|$ the conductance in the $N\gg1$ limit:
\begin{equation}
    \frac{G_\mathrm{2w}}{G_\infty} = \frac{1+e^{(N+1)\log \left(\frac{\nu_\mathrm{B}}{\nu_\mathrm{A}}\right)}}{1-e^{(N+1)\log \left(\frac{\nu_\mathrm{B}}{\nu_\mathrm{A}}\right)}} \nonumber= \coth\left[ \frac{N+1}{2} \log\left(\frac{\nu_\mathrm{A}}{\nu_\mathrm{B}}\right) \right], \label{eq_G_exponential}
\end{equation} 
which corresponds to Eq. (2) in the main text.
\section{Additional data for integer filling factors}
\subsection{Additional conductance data}
\begin{figure} [hbt!]
    \centering
    \includegraphics[width=1\linewidth]{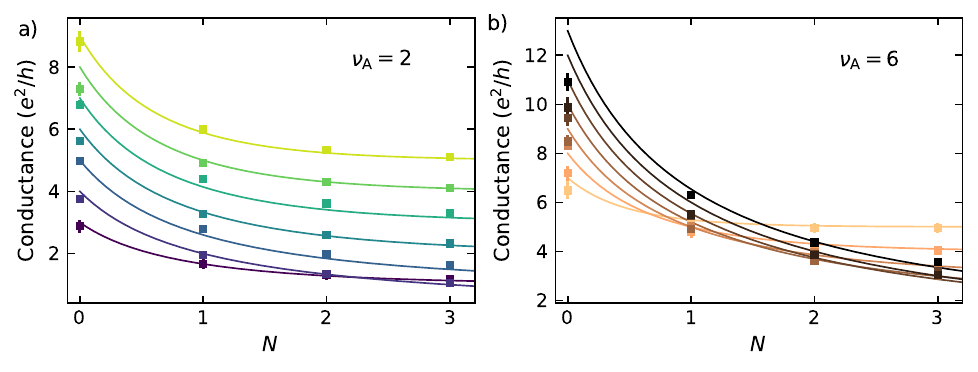}
    \caption{2-point conductance as a function of the number of intermediate Landauer reservoirs $N$, for (a) $\nu_\mathrm{A}=2$ and (b) $\nu_\mathrm{A}=6$. Color codes for different $(\nu_\mathrm{A},\nu_\mathrm{B})$ values are those used in the main text. Solid lines are the applications of Eq. \ref{eq_G}.}
    \label{SI_fig_G_all_raw} 
\end{figure}

In Fig. \ref{SI_fig_G_all_raw}, we present all the data in two-point conductance, as most were not shown in raw form in the main text.
Note that for $\nu_\mathrm{A}=6$, visible discrepancies between experimental points and theory (up to $15\,\%$) appear when all intermediate contacts are grounded, for conductances typically larger than $10G_Q$. This can be understood by taking into account the small but finite line resistance $\sim 1\,\mathrm{k}\Omega$ between the ground and the Landauer reservoir. We model simply the situation by writing current conservation for only one Landauer reservoir connected to ground via a resistance $r=1\,\mathrm{k}\Omega$ between source and drain. The correction $\delta G$ to the expected value $G_\Sigma=(\nu_\mathrm{A}+\nu_\mathrm{B})G_Q$ is:
\begin{equation}
    \delta G=-\frac{2\nu_\mathrm{A}\nu_\mathrm{B}}{\nu_\mathrm{A}+\nu_\mathrm{B}+h/re^2}.
\end{equation}
The maximum correction is expected for the (6,7) case, with $\delta G/G_\Sigma\approx -15\,\%$ as observed in our data.
\subsection{Individual filling factors}

Plateau regions of device A (resp. B) are identified by single gate sweep by setting device B (resp. A) at $\nu=0$.
\begin{figure}[hbt!]
    \centering
    \includegraphics[width=0.48\linewidth]{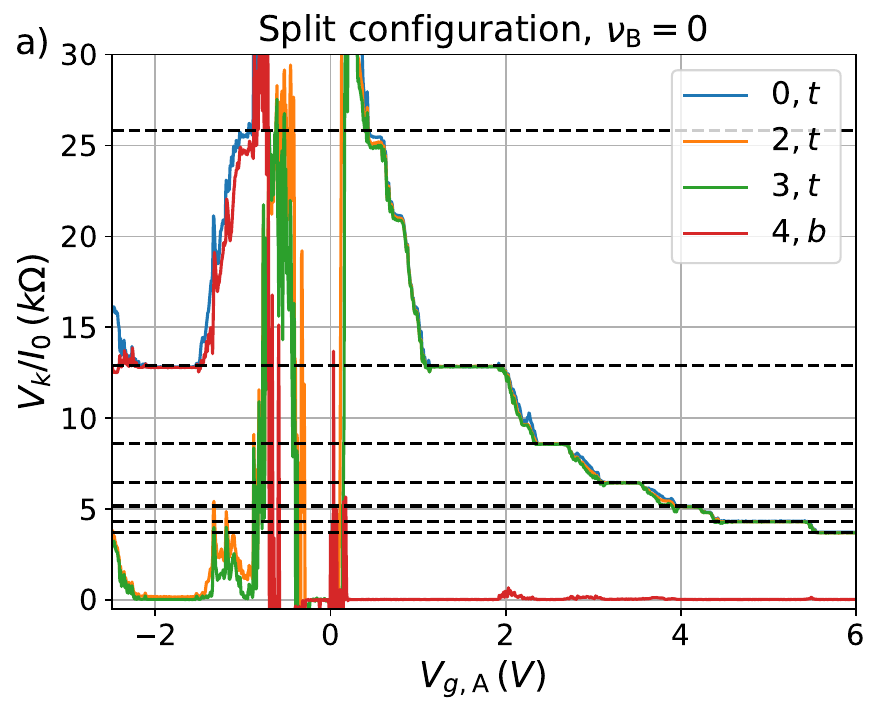}
    \includegraphics[width=0.48\linewidth]{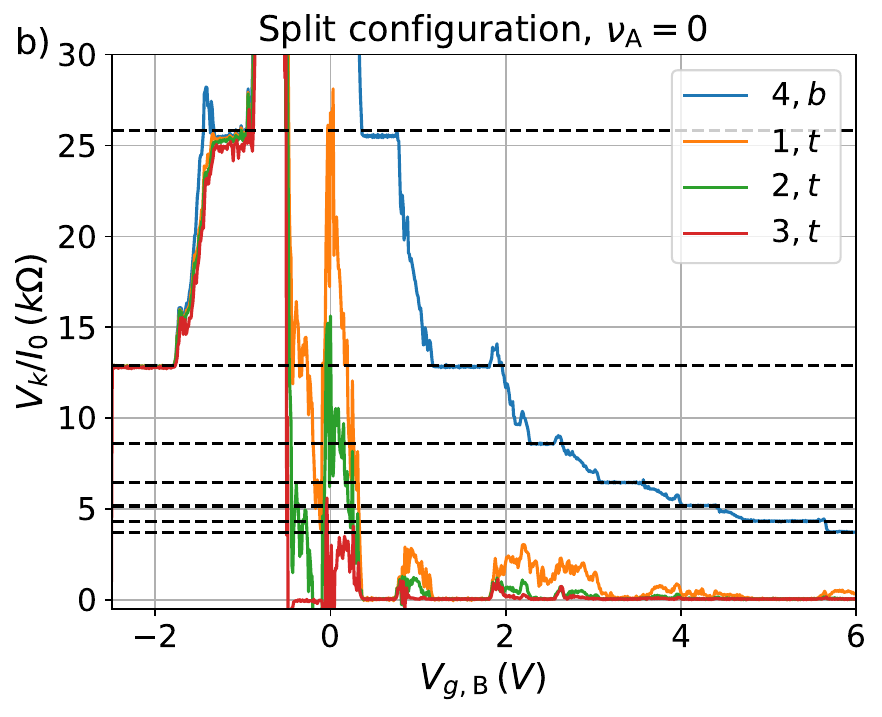}
    \caption{Gate sweep of Hall bar A (left panel) for $\nu_\mathrm{B}=0$ and B (right panel) for $\nu_\mathrm{A}=0$. Dashed horizontal lines show resistance plateaus $h/\nu e^2$, with $\nu=1...7$.}
    \label{SI_fig_sweep_nuA0_nuB_0} 
\end{figure}

\subsection{Gate/density sweep for $\nu_\mathrm{A}=6$}

In Fig 1 of the main text we show a sweep of device B's gate voltage/density (in filling factor units $\nu_\mathrm{B}$) while keeping $\nu_\mathrm{A}=2$, for top and bottom edge contacts. Here the equivalent graph is shown when keeping $\nu_\mathrm{A}=6$.

\begin{figure}[hbt!]
    \centering
    \includegraphics[width=0.6\linewidth]{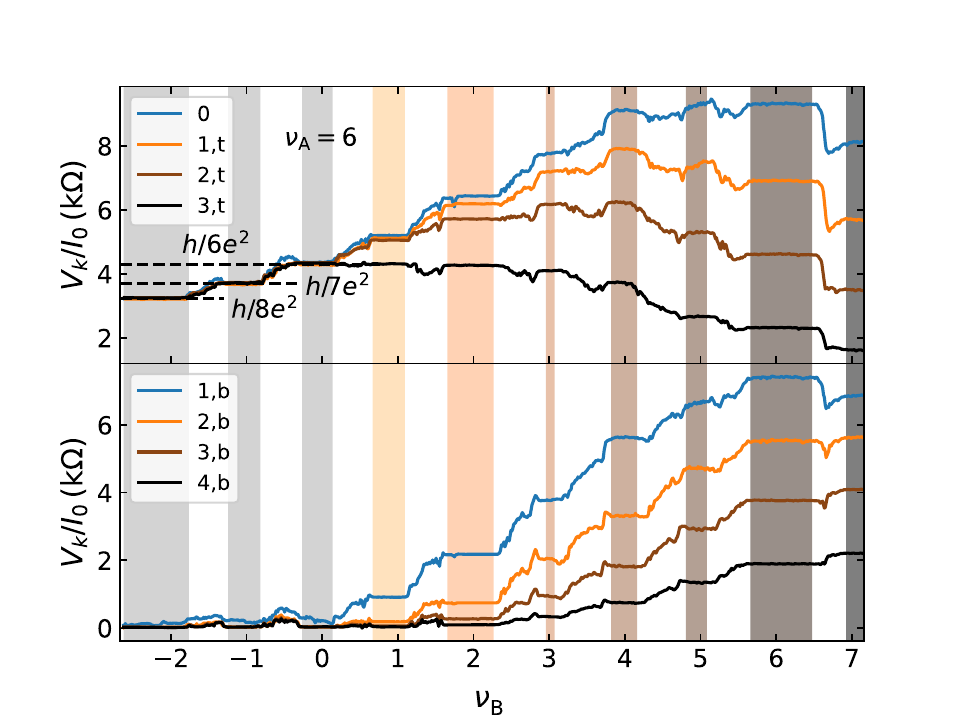}
    \caption{Gate sweep of Hall bar B for $\nu_\mathrm{A}=6$. The colors assigned to each plateau with $\nu_\mathrm{B}>0$ are those used in the main text.}
    \label{SI_fig_sweep_nuA6} 
\end{figure}

\section{Additional data}

\subsection{Additional data at $\nu_\mathrm{A}=1$, $\nu_\mathrm{B}=1/3$}

\begin{figure} [hbt!]
    \centering
    \includegraphics[width=0.48\linewidth]{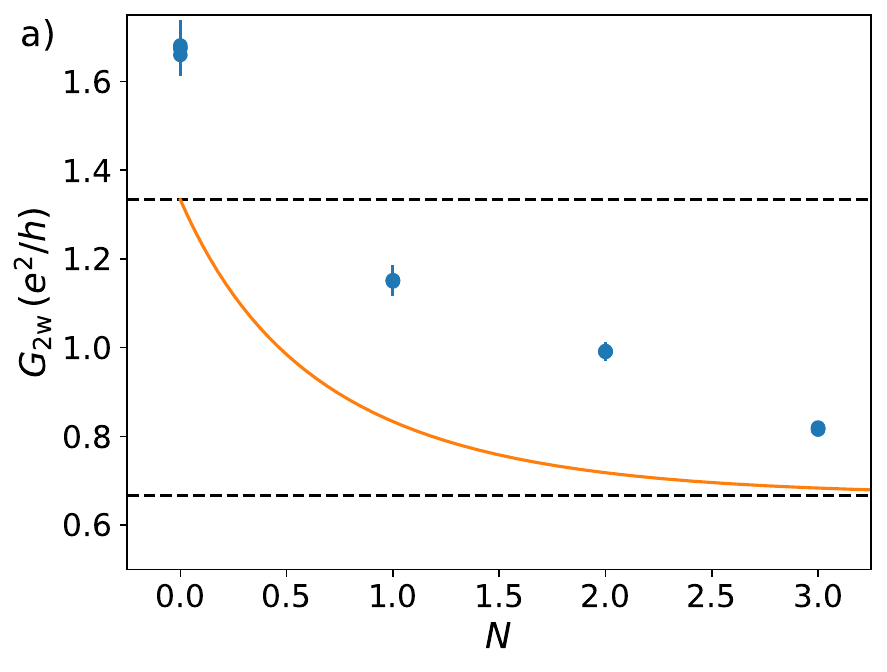}
    \includegraphics[width=0.48\linewidth]{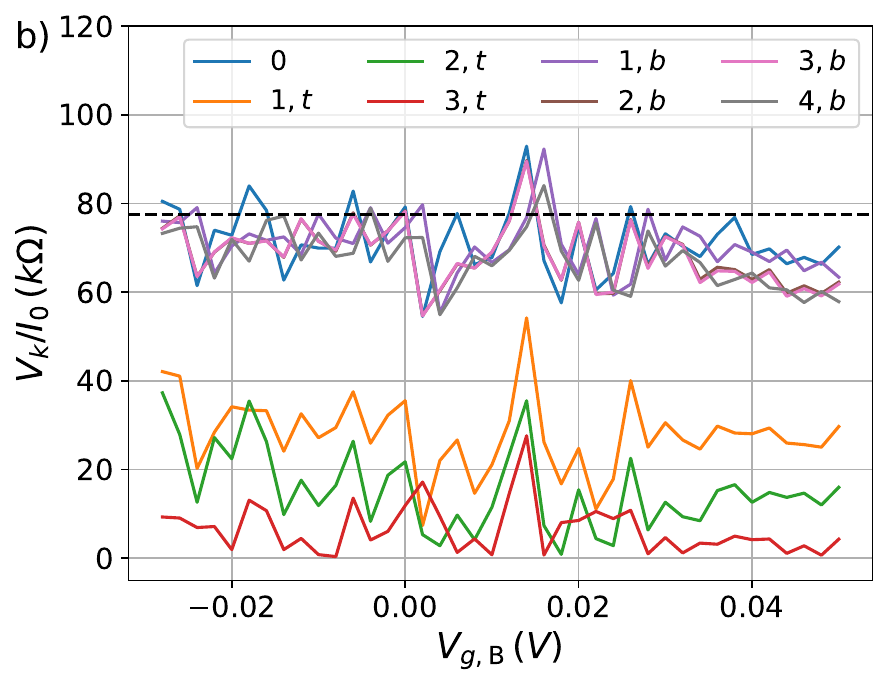}
    \caption{a) Two-point conductance as a function of the number of Landauer reservoirs $N$ on each edge for $\nu_\mathrm{A}=1$, $\nu_\mathrm{B}=1/3$. The solid line is the application of Eq. \ref{eq_G} for $\nu_A=1$, $\nu_B=1/3$, with the dashed lines marking the two limit values $G_\Sigma=4e^2/3h$ and $G_\infty=2e^2/3h$ expected theoretically. b) Voltages measured at each contact in the "Split" configuration, for $\nu_\mathrm{A}=0$, at $\nu_\mathrm{B}=1/3$, normalized to the input current.}
    \label{SI_G_frac} 
\end{figure}

Of particular interest is the 1-1/3 case, which emulates the widely investigated hole-conjugate Fractional Quantum Hall edge $\nu=2/3$ \cite{KFP_equilibration_1994,Cohen2019NComms,Lafont2019Science,SrivastavPRL2021,Hashisaka2021NComms,HashisakaPRX2023,Glattli2024arxiv}. There, charge conductance must converge exponentially to the equilibrated 2/3 value, while heat transport must be diffusive because of the equal heat conductances carried by the two counter-propagating edge channels. We present preliminary data in charge conductance, that do not agree with our simple theoretical model, as shown in Fig. \ref{SI_G_frac}. 

We attribute this discrepancy to significant backscattering at $\nu_\mathrm{A}=1/3$ for Hall bar A, as shown in Fig. \ref{SI_G_frac}b). Here leakage resistances up to $\sim 40\,\mathrm{k}\Omega$ are measured for t-type contacts contacts that are supposedly at zero voltage, while quantization at value $3h/e^2$ for is not well attained for b-type contacts. We stress that, as far as charge conductance is concerned, the theoretical expectation is completely equivalent, in terms of renormalized equilibration length, to the (6,2) and (2,6) cases presented here and in the main text.

\subsection{Additional data at $\nu_\mathrm{A}=2$, $\nu_\mathrm{B}=1/3,4/3$}

\begin{figure} [hbt!]
    \centering
    \includegraphics[width=0.48\linewidth]{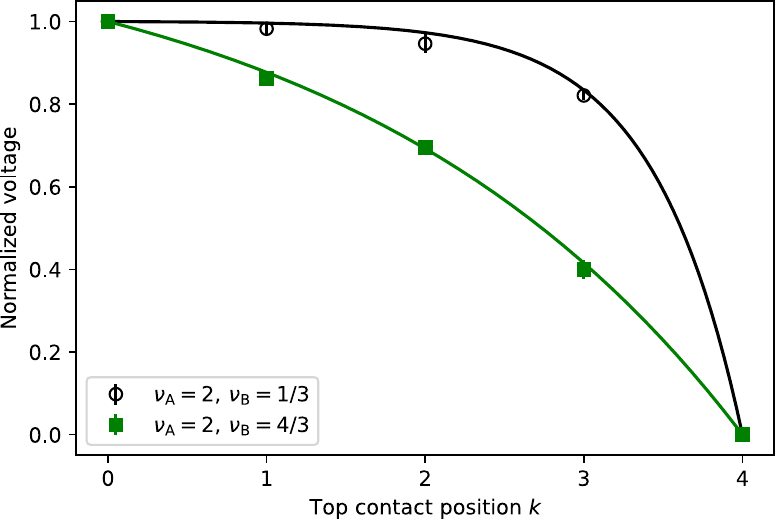}
    \includegraphics[width=0.48\linewidth]{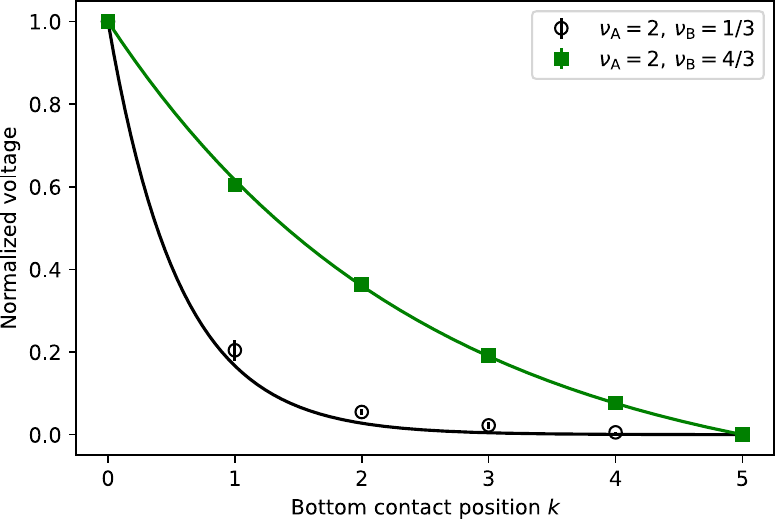}
    \caption{Normalized voltage profiles, in the "split" configuration on top (left) and bottom (right) contacts, for $\nu_\mathrm{A}=2$, at $\nu_\mathrm{B}=1/3$ and $4/3$. Lines are applications of Eq. \ref{eq_Vi_top_exp} (and its bottom equivalent), with corresponding filling factors.}
    \label{SI_Vprofile_frac} 
\end{figure}

In Figure \ref{SI_Vprofile_frac} we present preliminary data where the voltage profiles are measured for $\nu_\mathrm{A} = 2$ and two fractional filling factors $\nu_\mathrm{B} = 1/3,\,4/3$. Data are in good agreement with our theoretical model, despite the fact that it is in principle derived for non-interacting fermions. These states, similarly to the 1-1/3 case, present interest for charge-heat separation, which can be investigated with an improved setup featuring one connected edge, with the advantage provided by the $\nu=2$ plateau's robustness.
\section{Conductance in the incoherent regime for arbitrary $\nu_\mathrm{A,B}$}

In Ref. \cite{ProtopopovAnnals2017}, the authors obtain the two-point conductance in the vicinity of the KFP fixed point for the $\nu=2/3$ case, using a transfer matrix approach, which we briefly summarize and extend to arbitrary upstream and downstream edge conductances, in order to compare with the scattering approach used throughout our paper.

\begin{figure} [hbt!]
    \centering
    \includegraphics[width=0.75\linewidth]{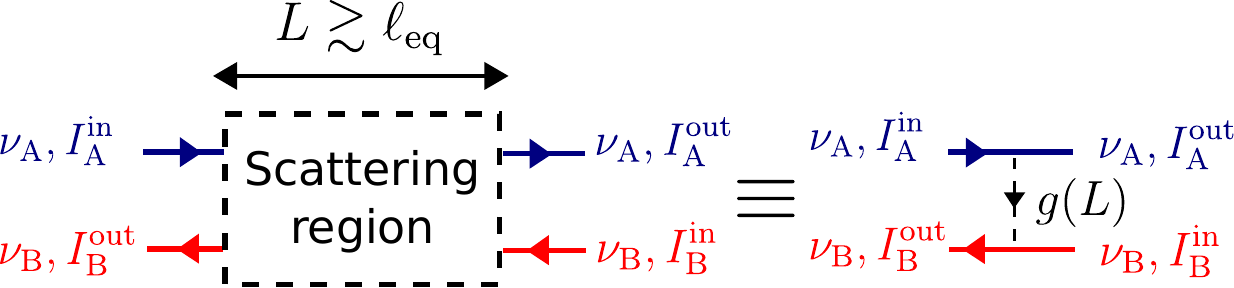}
    \caption{Scattering region along an edge hosting counter-propagating modes with conductances $\nu_\mathrm{A},\nu_\mathrm{B}$. Back-scattering over the region of size $L$ larger than the equilibration length is reduced to a single parameter, the edge-to-edge conductance $ge^2/h$.}
    \label{SI_transfer_interacting}
\end{figure}

The general transfer matrix for a scattering region may be parametrized by a single parameter, the edge-to-edge tunneling conductance $ge^2/h$, and writes:

\begin{equation}
\begin{pmatrix} 
    I^{\mathrm{out}}_\mathrm{A}\\ 
    I^{\mathrm{in}}_\mathrm{B} 
\end{pmatrix} =
\begin{pmatrix} 
   \frac{1-g(1/\nu_\mathrm{A}+1/\nu_\mathrm{B})}{1-g/\nu_\mathrm{B}} & \frac{g/\nu_\mathrm{B}}{1-g/\nu_\mathrm{B}}\\ 
    \frac{-g/\nu_\mathrm{A}}{1-g/\nu_\mathrm{B}}    & \frac{1}{1-g/\nu_\mathrm{B}}  
\end{pmatrix}
\begin{pmatrix} 
    I^{\mathrm{in}}_\mathrm{A}\\ 
    I^{\mathrm{out}}_\mathrm{B}
\end{pmatrix},
\end{equation}

where we have introduced the transfer matrix $\mathcal{T}_g$ connecting currents on the right of the scattering region to those on the left. To obtain the evolution of the tunneling conductance with the sample's length $L$, we consider two successive scattering regions (``subsegments") of lengths larger than the bare equilibration length $\ell_\mathrm{eq}$, hence being in the incoherent regime. By doing so, we can assume that no coherence subsists between subsegments 1 and 2. As a result, currents on the left of segment 1 can be linked to those on the right of segment 2 by a transfer matrix which is just the product of the two subsegments transfer matrices, $\mathcal{T}_g=\mathcal{T}_{g_1}\mathcal{T}_{g_2}$. Here, the effective tunneling conductance $g$ is linked to the subsegments tunneling conductances via the relation:

\begin{equation}
    \label{RG_transfo_g}
    g=\frac{g_1+g_2-g_1g_2(1/\nu_\mathrm{A}+1/\nu_\mathrm{B})}{1-g_1g_2/\nu_\mathrm{A}\nu_\mathrm{B}}.
\end{equation}

By iterating the operation for larger and larger segments, we eventually reach the fixed point limit where $g=g_1=g_2$, with the one physical solution for Eq. (\ref{RG_transfo_g}), $g=\min(\nu_\mathrm{A},\nu_\mathrm{B})$. Of interest is the evolution equation of $g$ in the vicinity of the fixed point, with the sample's reduced length $y\sim L/\ell_\mathrm{eq}$:

\begin{equation}
    \label{RG_evolution_g_fixP}
    \frac{\mathrm{d}g}{\mathrm{d}y}=1-g(1/\nu_\mathrm{A}+1/\nu_\mathrm{B})+g^2/\nu_\mathrm{A}\nu_\mathrm{B}.
\end{equation}

For $\nu_\mathrm{A}\neq\nu_\mathrm{B}$ Eq. (\ref{RG_evolution_g_fixP}) can be solved, yielding exponential convergence to the fixed point value:
\begin{equation}
    \label{g_incoh_exp}
    g=\nu_\mathrm{B}\frac{1+\frac{\nu_\mathrm{A}}{\nu_\mathrm{B}}e^{-L/\tilde{\ell}_\mathrm{eq}+K(\nu_\mathrm{A}-\nu_\mathrm{B})}}{1-e^{-L/\tilde{\ell}_\mathrm{eq}+K(\nu_\mathrm{A}-\nu_\mathrm{B})}},
\end{equation}

where $K$ is a constant that depends on the UV cutoff (that is, the initial smallest length $\ell_\mathrm{eq}$ in the iteration procedure), and where we introduce the \textit{effective, filling factor-dependent} equilibration length:

\begin{equation}
    \label{effective_leq}
    \tilde{\ell}_\mathrm{eq}=\frac{\ell_\mathrm{eq}}{1/\nu_\mathrm{B}-1/\nu_\mathrm{A}},
\end{equation}

where we recover the expression found in previous works \cite{Lin2019PRB,SpanslattPRB2020}. We stress that in this section $\tilde{\ell}_\mathrm{eq}$ differs from the one introduced in the main text. This is due to the fundamentally different backscattering mechanisms involved (tunneling element vs. Landauer reservoir). However, as $\nu_\mathrm{A}\rightarrow\nu_\mathrm{B}$, the two versions present the same divergence $\tilde{\ell}_\mathrm{eq}\sim\ell_\mathrm{eq}/\epsilon$, with $\epsilon=|\nu_\mathrm{A}-\nu_\mathrm{B}|$.

For the critical case $\nu_\mathrm{A}=\nu_\mathrm{B}=\nu$, we obtain an algebraic decay of the tunneling conductance towards its fixed point $\nu$:

\begin{equation}
\label{g_incoh_alg}
    g(y)=\frac{\nu y-\nu^2+\alpha\nu^3}{y+\alpha\nu^2},
\end{equation}

with $\alpha\sim 1$ a UV cutoff-dependent constant.

We then proceed to write the two-point conductance, as a function of the tunneling conductances on both edges:

\begin{equation}
    \label{2pt_g_generic}
G_\mathrm{2w}(L)=e^2/h[\nu_\mathrm{A}+\nu_\mathrm{B}-g_\mathrm{t}(L)-g_\mathrm{b}(L)].
\end{equation}

Without loss of generality, we consider two edges that have the same length. For $\nu_\mathrm{A}>\nu_\mathrm{B}$, we obtain:

\begin{equation}
    \label{2pt_g_unequal}
G_\mathrm{2w}(L)=\frac{(\nu_\mathrm{A}-\nu_\mathrm{B})e^2}{h}\coth{\left[L/2\tilde{\ell}_\mathrm{eq}+K\ell_\mathrm{eq}/\tilde{\ell}_\mathrm{eq}\right]}.
\end{equation}

Hence conductance gets exponentially close to its equilibrated value for $L\gg\ell_\mathrm{eq}$. Note that, once the filling factor dependence is absorbed in $\tilde{\ell}_\mathrm{eq}$, this expression is equivalent to Eq. (2) of the main text, with the discrete-to-continuum mapping $N=L/\ell_\mathrm{eq}$ and $K=1/2$ (recall that in the main text $\delta=\tilde{\ell}_\mathrm{eq}/\ell_\mathrm{eq}$). For $\nu_\mathrm{A}=\nu_\mathrm{B}=\nu$, on the other hand, we obtain an Ohmic scaling of the conductance, signaling a diffusive regime, or equivalently, a diverging effective equilibration length $\tilde{\ell}_\mathrm{eq}$:

\begin{equation}
    \label{2pt_g_equal}
    G_\mathrm{2w}=\frac{2\nu^2e^2/h}{y+\alpha\nu^2}\underset{L\gg\ell_{\mathrm{eq}}}{\sim}\frac{2\nu^2e^2\ell_\mathrm{eq}}{hL},
\end{equation}

which again follows the same trend as the critical regime highlighted in the main text for $N=L/\ell_\mathrm{eq}\gg 1$. This can be seen as a signature of scale invariance: in the critical regime, the conductance follows an Ohmic scaling irrespective of the microscopic equilibration mechanisms.

Note that the physics investigated there, and its emulation in our system (by construction since adding one reservoir = adding one $\ell_\mathrm{eq}$), is valid only for length scales larger than the equilibration length (otherwise $\mathcal{T}_g\neq\mathcal{T}_{g_1}\mathcal{T}_{g_2}$). It neglects interference effects that occur at smaller scale and give rise to a disorder-dominated regime \cite{HashisakaPRX2023} in the strong coupling phase of the KFP diagram \cite{KFP_equilibration_1994}.


\end{document}